\documentclass[12pt]{article}
\usepackage{euler}
\usepackage{amsmath}
\usepackage{amssymb}
\usepackage{amstext}
\usepackage{amscd}
\usepackage{amsfonts}
\DeclareMathAlphabet{\EuFrak}{U}{euf}{m}{n}
\DeclareMathAlphabet{\EuScript}{U}{eus}{m}{n}

\title{{\bf Convolution of Ultradistributions,
Field Theory, Lorentz Invariance and Resonances}
\thanks{\it{This work was partially supported by Consejo
Nacional
de Investigaciones Cient\'{\i}ficas and Comisi\'{o}n de
Investigaciones Cient\'{\i}ficas de la Pcia. de Buenos
Aires;
Argentina.}}}
\author{C.G.Bollini, P. Marchiano and M.C.Rocca\\
Departamento de F\'{\i}sica, Fac. de Ciencias Exactas,\\
Universidad Nacional de La Plata.\\
C.C. 67 (1900) La Plata. Argentina.}

\date{May 1, 2006}

\begin{document}

\maketitle

\begin{abstract}

In this work, a general definition of convolution between two
arbitrary  Ultradistributions of Exponential type (UET) is given.
The product of two arbitrary  UET is defined via the convolution of its
corresponding Fourier Transforms.
Some  examples of convolution of two UET are given.

Expressions for the Fourier Transform of
spherically symmetric (in Euclidean space)
and Lorentz invariant (in Minkowskian space)
UET  in term of modified Bessel distributions are obtained
(Generalization of Bochner's theorem).

The generalization to UET of dimensional regularization in
configuration space is obtained in both, Euclidean and
Minkowskian spaces

As an application of our formalism, we give
a solution to the question of normalization of
resonances in Quantum Mechanics.

General formulae for convolution of even,
spherically symmetric and Lorentz invariant UET
are obtained and several examples
of application are given.


\end{abstract}

\newpage

\renewcommand{\theequation}{\arabic{section}.\arabic{equation}}

\section{Introduction}

The question of the product of distributions with coincident point
singularities is related in Field Theory, to the asymptotic
behavior of loop integrals of propagators.
From a mathematical point of view,
the question reduces to define a product in a ring with
zero-factors.
As is known, the usual definitions lead
to limitations on the set of distributions that can be multiplied
together to give another distribution of the same kind.
The properties of tempered ultradistributions (ref.\cite{tp1,tp2}) are well
adapted for their use in Field Theory. In this respect we have
shown (ref.\cite{tp3,tp4,tp5}) that it is possible to define
the convolution of any pair of tempered
Ultradistributions, giving as a result another tempered
Ultradistribution.

Ultradistributions also have the
advantage of being representable by means of analytic functions.
So that, in general, they are easier to work with them and,
as we shall see, have interesting properties. One of those properties
is that Schwartz's tempered distributions are canonical and continuously
injected into tempered Ultradistributions
and as a consequence the Rigged
Hilbert Space  with tempered distributions is  canonical and continuously
included
in the Rigged Hilbert Space with tempered Ultradistributions.

A further step is to consider Ultradistributions of Exponential Type
(ref.\cite{tp1,tp6}),
and to define a convolution product between any pair of them.
This is made in this paper together some examples and applications.
In this case Schwartz's tempered dsitributions and Sebastiao e Silva's
tempered Ultradistributions are canonical and continuously
injected
into Ultradistributions of
Exponential Type
and as a consequence the Rigged
Hilbert Spaces  with tempered distributions and
tempered Ultradistributions  are  canonical and continuously
included
in the Rigged Hilbert Space with Ultradistributions of Exponential Type.

Furthermore, Ultradistributions of Exponential Type are adequates to describe
Gamow States and exponentially increasing fields in Quantun
Field Theory as we show in this paper.

This paper is organized as follow:
in section 2  we define the Ultradistributions of Exponential Type
and their Fourier transform. They
are  part of a Guelfand's Triplet ( or Rigged Hilbert Space \cite{tp7} )
together with their respective duals and a ``middle term'' Hilbert
space.
In section 3 we give a general expression for the Fourier
transform of a spherically symmetric Ultraditributions
of Exponential Type (UET) and some
examples ot it.
In section 4 we obtain  the expression for the
Fourier transform of Lorentz invariant UET and
we give some examples of their use.
In section 5 we give the generalization to
UET  of  ``dimensional
regularization in configuration space'' obtained in ref.\cite{tp5,tp8}.
In section 6 we give the general formula for the convolution
of two UET  and some
examples.
In section 7 we give a solution to the question of normalization
of Gamow states. We give as example the l=0 state of
Square-Well potential.
In section 8 we present the formula for the convolution of
two even UET, and
various examples.
In section 9, we give the general formula for the convolution of
two Spherically Symmetric UET  and several examples.
In section 10 we treat the final topic of this paper:
the formula for the convolution
of two Lorentz Invariant Ultradistributions
of Exponential Type. We obtain it
and give examples of their use.
Finally, section 11  is reserved for a discussion of the principal results.

\section{Ultradistributions of Exponential Type}

Let ${\cal S}$ be the space of Schwartz of test functions rapidly decreasing.
Let ${\Lambda}_j$ be the region of the complex plane defined as:
\begin{equation}
\label{er2.1}
{\Lambda}_j=\left\{z\in\boldsymbol{\mathbb{C}} :
|\Im(z)|< j : j\in\boldsymbol{\mathbb{N}}\right\}
\end{equation}
According to ref.\cite{tp1,tp6} the space of test functions $\hat{\phi}\in
{\mathcal{\large{V}}}_j$ is
constituted by all entire analytic functions of ${\cal S}$ for which
\begin{equation}
\label{ep2.2}
||\hat{\phi} ||_j=\max_{k\leq j}\left\{\sup_{z\in{\Lambda}_j}\left[e^{(j|\Re (z)|)}
|{\hat{\phi}}^{(k)}(z)|\right]\right\}
\end{equation}
is finite.\\
The space $\mathcal{\large{Z}}$ is then defined as:
\begin{equation}
\label{er2.3}
\mathcal{\large{Z}} =\bigcap_{j=0}^{\infty} {\mathcal{\large{V}}}_j
\end{equation}
It is a complete countably normed space with the topology generated by
the system of semi-norms $\{||\cdot ||_j\}_{j\in \mathbb{N}}$.
The dual of $\mathcal{\large{Z}}$, denoted by
$\mathcal{\large{B}}$, is by definition the space of ultradistributions of exponential
type (ref.\cite{tp1,tp6}).
Let $\mathcal{S}$ the space of rapidly decreasing sequences. According to
ref.\cite{tp7} $\mathcal{S}$ is a nuclear space. We consider now the space of
sequences $\mathcal{P}$ generated by the Taylor development of
$\hat{\phi}\in\mathcal{\large{Z}}$
\begin{equation}
\label{er2.4}
\mathcal{P}=\left\{\mathcal{Q} : \mathcal{Q}
\left(\hat{\phi}(0),{\hat{\phi}}^{'}(0),\frac {{\hat{\phi}}^{''}(0)} {2},...,
\frac {{\hat{\phi}}^{(n)}(0)} {n!},...\right) : \hat{\phi}\in\mathcal{Z}\right\}
\end{equation}
The norms that define the topology of $\mathcal{P}$ are given by:
\begin{equation}
\label{er2.5}
||\hat{\phi} ||^{'}_p=\sup_n \frac {n^p} {n} |{\hat{\phi}}^n(0)|
\end{equation}
$\mathcal{P}$ is a subespace of $\mathcal{S}$ and therefore is a nuclear space.
As the norms $||\cdot ||_j$ and $||\cdot ||^{'}_p$ are equivalent, the correspondence
\begin{equation}
\label{er2.6}
\mathcal{\large{Z}}\Longleftrightarrow \mathcal{P}
\end{equation}
is an isomorphism and therefore $\mathcal{Z}$ is a countably normed nuclear space.
We can define now the set of scalar products
\[<\hat{\phi}(z),\hat{\psi}(z)>_n=\sum\limits_{q=0}^n\int\limits_{-\infty}^{\infty}e^{2n|z|}
\overline{{\hat{\phi}}^{(q)}}(z){\hat{\psi}}^{(q)}(z)\;dz=\]
\begin{equation}
\label{er2.7}
\sum\limits_{q=0}^n\int\limits_{-\infty}^{\infty}e^{2n|x|}
\overline{{\hat{\phi}}^{(q)}}(x){\hat{\psi}}^{(q)}(x)\;dx
\end{equation}
This scalar product induces the norm
\begin{equation}
\label{er2.8}
||\hat{\phi}||_n^{''}=[<\hat{\phi}(x),\hat{\phi}(x)>_n]^{\frac {1} {2}}
\end{equation}
The norms $||\cdot ||_j$ and $||\cdot ||^{''}_n$ are equivalent, and therefore
$\mathcal{\large{Z}}$ is a countably hilbertian nuclear space.
Thus, if we call now ${\mathcal{\large{Z}}}_p$ the completion of
$\mathcal{\large{Z}}$ by the norm $p$ given in (\ref{er2.8}), we have:
\begin{equation}
\label{er2.9}
\mathcal{\large{Z}}=\bigcap_{p=0}^{\infty}{\mathcal{\large{Z}}}_p
\end{equation}
where
\begin{equation}
\label{er2.10}
{\mathcal{\large{Z}}}_0=\boldsymbol{H}
\end{equation}
is the Hilbert space of square integrable functions.\\
As a consequence the ``nested space''
\begin{equation}
\label{er2.11}
\mathcal{\Large{U}}=\boldsymbol{(}\mathcal{\large{Z}},
\boldsymbol{H}, \mathcal{\large{B}}\boldsymbol{)}
\end{equation}
is a Guelfand's triplet (or a Rigged Hilbert space=RHS. See ref.\cite{tp7}).

Any Guelfand's triplet
$\mathcal{\Large{G}}=\boldsymbol{(}\boldsymbol{\Phi},
\boldsymbol{H},\boldsymbol{{\Phi}^{'}}\boldsymbol{)}$
has the fundamental property that a linear and symmetric operator
on $\boldsymbol{\Phi}$, admitting an extension to a self-adjoint
operator in
$\boldsymbol{H}$, has a complete set of generalized eigen-functions
in $\boldsymbol{{\Phi}^{'}}$ with real eigenvalues.

$\mathcal{\large{B}}$ can also be characterized in the following way
( refs.\cite{tp1},\cite{tp6} ): let ${\mathcal{E}}_{\omega}$ be the space of
all functions $\hat{F}(z)$ such that:

${\Large {\boldsymbol{I}}}$-
$\hat{F}(z)$ is analytic for $\{z\in \boldsymbol{\mathbb{C}} :
|Im(z)|>p\}$.

${\Large {\boldsymbol{II}}}$-
$\hat{F}(z)e^{-p|\Re(z)|}/z^p$ is bounded continuous  in
$\{z\in \boldsymbol{\mathbb{C}} :|Im(z)|\geqq p\}$,
where $p=0,1,2,...$ depends on $\hat{F}(z)$.

Let $\mathcal{N}$ be:
$\mathcal{N}=\{\hat{F}(z)\in{\mathcal{E}}_{\omega} :\hat{F}(z)\; \rm{is\; entire\; analytic}\}$.
Then $\mathcal{\large{B}}$ is the quotient space:

${\Large {\boldsymbol{III}}}$-
$\mathcal{\large{B}}={\mathcal{E}}_{\omega}/\mathcal{N}$

Due to these properties it is possible to represent any ultradistribution
as ( ref.\cite{tp1,tp6} ):
\begin{equation}
\label{er2.12}
\hat{F}(\hat{\phi})=<\hat{F}(z), \hat{\phi}(z)>=\oint\limits_{\Gamma} \hat{F}(z) \hat{\phi}(z)\;dz
\end{equation}
where the path ${\Gamma}_j$ runs parallel to the real axis from
$-\infty$ to $\infty$ for $Im(z)>\zeta$, $\zeta>p$ and back from
$\infty$ to $-\infty$ for $Im(z)<-\zeta$, $-\zeta<-p$.
( $\Gamma$ surrounds all the singularities of $\hat{F}(z)$ ).

Formula (\ref{er2.12}) will be our fundamental representation for a tempered
ultradistribution. Sometimes use will be made of ``Dirac formula''
for exponential ultradistributions ( ref.\cite{tp1} ):
\begin{equation}
\label{er2.13}
\hat{F}(z)\equiv\frac {1} {2\pi i}\int\limits_{-\infty}^{\infty}
\frac {\hat{f}(t)} {t-z}\;dt\equiv
\frac {\cosh(\lambda z)} {2\pi i}\int\limits_{-\infty}^{\infty}
\frac {\hat{f}(t)} {(t-z)\cosh(\lambda t)}\;dt
\end{equation}
where the ``density'' $\hat{f}(t)$ is such that
\begin{equation}
\label{er2.14}
\oint\limits_{\Gamma} \hat{F}(z) \hat{\phi}(z)\;dz =
\int\limits_{-\infty}^{\infty} \hat{f}(t) \hat{\phi}(t)\;dt
\end{equation}
(\ref{er2.13}) should be used carefully in this case.
While $\hat{F}(z)$ is analytic on $\Gamma$, the density $\hat{f}(t)$ is in
general singular, so that the r.h.s. of (\ref{er2.14}) should be interpreted
in the sense of distribution theory.

Another important property of the analytic representation is the fact
that on $\Gamma$, $\hat{F}(z)$ is bounded by a exponential and a power of $z$
( ref.\cite{tp1,tp6} ):
\begin{equation}
\label{er2.15}
|\hat{F}(z)|\leq C|z|^pe^{p|\Re(z)|}
\end{equation}
where $C$ and $p$ depend on $\hat{F}$.

The representation (\ref{er2.12}) implies that the addition of any entire function
$\hat{G}(z\in\mathcal{N})$ to $\hat{F}(z)$ does not alter the ultradistribution:
\[\oint\limits_{\Gamma}\{\hat{F}(z)+\hat{G}(z)\}\hat{\phi}(z)\;dz=
\oint\limits_{\Gamma} \hat{F}(z)\hat{\phi}(z)\;dz+\oint\limits_{\Gamma}
\hat{G}(z)\hat{\phi}(z)\;dz\]
But:
\[\oint\limits_{\Gamma} \hat{G}(z)\hat{\phi}(z)\;dz=0\]
as $\hat{G}(z)\hat{\phi}(z)$ is entire analytic
( and rapidly decreasing ),
\begin{equation}
\label{er2.16}
\therefore \;\;\;\;\oint\limits_{\Gamma} \{\hat{F}(z)+\hat{G}(z)\}\hat{\phi}(z)\;dz=
\oint\limits_{\Gamma} \hat{F}(z)\hat{\phi}(z)\;dz
\end{equation}

Another very important property of $\mathcal{\large{B}}$ is that
$\mathcal{\large{B}}$ is reflexive under the Fourier transform:
\begin{equation}
\label{er2.17}
\mathcal{\large{B}}={\cal F}_c\left\{\mathcal{\large{B}}\right\}=
{\cal F}\left\{\mathcal{\large{B}}\right\}
\end{equation}
where the complex Fourier transform $F(k)$ of $\hat{F}(z)\in\mathcal{\large{B}}$
is given by:
\[F(k)=\Theta[\Im(k)]\int\limits_{{\Gamma}_+}\hat{F}(z)e^{ikz}\;dz-
\Theta[-\Im(k)]\int\limits_{{\Gamma}_{-}}\hat{F}(z)e^{ikz}\;dz=\]
\begin{equation}
\label{er2.18}
\Theta[\Im(k)]\int\limits_0^{\infty}\hat{f}(x)e^{ikx}\;dx-
\Theta[-\Im(k)]\int\limits_{-\infty}^0\hat{f}(x) e^{ikx}\;dx
\end{equation}
Here ${\Gamma}_+$ is the part of $\Gamma$ with $\Re(z)\geq 0$ and
${\Gamma}_{-}$ is the part of $\Gamma$ with $\Re(z)\leq 0$
Using (\ref{er2.18}) we can interpret Dirac's formula as:
\begin{equation}
\label{er2.19}
F(k)\equiv\frac {1} {2\pi i}\int\limits_{-\infty}^{\infty}
\frac {f(s)} {s-k}\; ds\equiv{\cal F}_c\left\{{\cal F}^{-1}\left\{f(s)\right\}\right\}
\end{equation}
The treatment for ultradistributions of exponential type defined on
${\boldsymbol{\mathbb{C}}}^n$ is similar to the one-variable.
In this case
\begin{equation}
\label{er2.20}
{\Lambda}_j=\left\{z=(z_1, z_2,...,z_n)\in{\boldsymbol{\mathbb{C}}}^n :
|\Im(z_k)|\leq j\;\;\;1\leq k\leq n\right\}
\end{equation}
\begin{equation}
\label{er2.21}
||\hat{\phi} ||_j=\max_{k\leq j}\left\{\sup_{z\in{\Lambda}_j}\left[
e^{j\left[\sum\limits_{p=1}^n|\Re(z_p)|\right]}\left| D^{(k)}\hat{\phi}(z)\right|\right]\right\}
\end{equation}
where $D^{(k)}={\partial}^{(k_1)}{\partial}^{(k_2)}\cdot\cdot\cdot{\partial}^{(k_n)}\;\;\;\;
k=k_1+k_2+\cdot\cdot\cdot+k_n$

$\mathcal{\large{B}}$ is characterized as follows. Let
${\mathcal{E}}_{\omega}$ be the space of all functions $\hat{F}(z)$ such that:

${\Large {\boldsymbol{I}}}^{'}$-
$\hat{F}(z)$ is analytic for $\{z\in \boldsymbol{{\mathbb{C}}^n} :
|Im(z_1)|>p, |Im(z_2)|>p,...,|Im(z_n)|>p\}$.

${\Large {\boldsymbol{II}}}^{'}$-
$\hat{F}(z)e^{-\left[p\sum\limits_{j=1}^n|\Re(z_j)|\right]}/z^p$
is bounded continuous  in
$\{z\in \boldsymbol{{\mathbb{C}}^n} :|Im(z_1)|\geqq p,|Im(z_2)|\geqq p,
...,|Im(z_n)|\geqq p\}$,
where $p=0,1,2,...$ depends on $\hat{F}(z)$.

Let $\mathcal{N}$ be:
$\mathcal{N}=\left\{\hat{F}(z)\in{\mathcal{E}}_{\omega} :\hat{F}(z)\;\right.$
is entire analytic at minus in one of the variables $\left. z_j\;\;\;1\leq j\leq n\right\}$
Then $\mathcal{\large{B}}$ is the quotient space:

${\Large {\boldsymbol{III}}}^{'}$-
$\mathcal{\large{B}}={\mathcal{E}}_{\omega}/\mathcal{N}$
We have now
\begin{equation}
\label{er2.22}
\hat{F}(\hat{\phi})=<\hat{F}(z), \hat{\phi}(z)>=\oint\limits_{\Gamma} \hat{F}(z) \hat{\phi}(z)\;
dz_1\;dz_2\cdot\cdot\cdot dz_n
\end{equation}
$\Gamma={\Gamma}_1\cup{\Gamma}_2\cup ...{\Gamma}_n$
where the path ${\Gamma}_j$ runs parallel to the real axis from
$-\infty$ to $\infty$ for $Im(z_j)>\zeta$, $\zeta>p$ and back from
$\infty$ to $-\infty$ for $Im(z_j)<-\zeta$, $-\zeta<-p$.
(Again $\Gamma$ surrounds all the singularities of $\hat{F}(z)$ ).
The n-dimensional Dirac's formula is
\begin{equation}
\label{ep2.23}
\hat{F}(z)=\frac {1} {(2\pi i)^n}\int\limits_{-\infty}^{\infty}
\frac {\hat{f}(t)} {(t_1-z_1)(t_2-z_2)...(t_n-z_n)}\;dt_1\;dt_2\cdot\cdot\cdot dt_n
\end{equation}
where the ``density'' $\hat{f}(t)$ is such that
\begin{equation}
\label{ep2.24}
\oint\limits_{\Gamma} \hat{F}(z)\hat{\phi}(z)\;dz_1\;dz_2\cdot\cdot\cdot dz_n =
\int\limits_{-\infty}^{\infty} f(t) \hat{\phi}(t)\;dt_1\;dt_2\cdot\cdot\cdot dt_n
\end{equation}
and the modulus of $\hat{F}(z)$ is bounded by
\begin{equation}
\label{er2.25}
|\hat{F}(z)|\leq C|z|^p e^{\left[p\sum\limits_{j=1}^n|\Re(z_j)|\right]}
\end{equation}
where $C$ and $p$ depend on $\hat{F}$.

\section{The Fourier Transform in Euclidean Space}

\setcounter{equation}{0}

We define a spherically symmetric ultradistribution of exponential type $\hat{F}(z)$
as a ultradistribution of exponential type such that $\hat{f}(t)$ in (\ref{ep2.23}) is
spherically symmetric (Note that a spherically symmetric
ultradistribution is not  necessarily spherically symmetric in an explicit way).
In this case we can use for the Fourier transform of $\hat{f}(t)$
the formula obtained in ref.\cite{tp5}:
\[F(\rho)=\frac {(2\pi)^{\frac {\nu-2} {2}}} {{\rho}^{\frac {\nu-2} {4}}}
\int\limits_0^{\infty}\hat{f}(x)x^{\frac {\nu-2} {4}}\left\{\Theta[\Im(\rho)]
e^{-\frac {i\pi\nu} {4}}{\cal{K}}_{\frac {\nu-2} {2}}(-ix^{1/2}{\rho}^{1/2})-\right.\]
\[\left.\Theta[-\Im(\rho)]
e^{\frac {i\pi\nu} {4}}{\cal{K}}_{\frac {\nu-2} {2}}(ix^{1/2}{\rho}^{1/2})\right\}\;dx\; +\]
\begin{equation}
\label{ep3.1}
\frac {2{\pi}^{\frac {\nu-2} {2}}} {\Gamma(\frac {\nu-2} {2})
{\rho}^{\frac {\nu-2} {4}}}
\int\limits_0^{\infty}\hat{f}(x)x^{\frac {\nu-2} {4}} {\cal{S}}_{\frac {\nu-4} {2},
\frac {\nu-2} {2}}(x^{1/2}{\rho}^{1/2})\;dx
\end{equation}
When $\nu=2n$, n an integer number
${\rho}^{\frac {2-\nu} {4}} {\cal {S}}_{\frac {\nu-4} {2},\frac {\nu-2} {2}}$
is null. In fact
\begin{equation}
\label{ep3.2}
{\rho}^{\frac {2-\nu} {4}} {\cal {S}}_{\frac {\nu-4} {2},\frac {\nu-2} {2}}=
\sum\limits_{m=0}^{\frac {\nu-4} {2}}\frac {(\frac {\nu} {2}-m)!} {m!}
4^{\frac {\nu-2-4m} {4}} x^{\frac {4m+2-\nu} {4}}{\rho}^{\frac {2m+2-\nu} {2}}=0
\end{equation}
that is null in the complex variable $\rho$ in a space of dimension $\nu=2n$.
Thus in this case the second integral in (\ref{ep3.1}) vanishes and it becomes in:
\[F(\rho)=\frac {(2\pi)^{\frac {\nu-2} {2}}} {\rho^{\frac {\nu-2} {4}}}\int\limits_0^{\infty}\hat{f}(x)
x^{\frac {\nu-2} {4}}\left[\Theta[\Im(\rho)]
e^{-i\frac {\pi} {4}\nu}\boldsymbol{\cal{K}}_{\frac {\nu-2} {2}}(-ix^{1/2}{\rho}^{1/2})\right.\]
\begin{equation}
\label{ep3.3}
\left. -\Theta[-\Im(\rho)]
 e^{i\frac {\pi} {4}\nu}\boldsymbol{\cal{K}}_{\frac {\nu-2} {2}}(ix^{1/2}{\rho}^{1/2})\right]\;dx
\end{equation}
In the next section we shall see that formulae  (\ref{ep3.2}), (\ref{ep3.3})
can be generalized to Minkowskian space.

When $\hat{f}$ is a spherically symmetric , we can use
(\ref{ep3.3}) to define its Fourier transform. In addition for $\nu=2n$ we can follow
the treatment of ref.\cite{tp9} to define the Fourier transform. Thus we have
\begin{equation}
\label{ep3.4}
\int\limits_{-\infty}^{\infty}f(\rho)\phi(\rho){\rho}^{\frac {\nu-2} {2}}\;d\rho=(2\pi)^{\nu}
\int\limits_0^{\infty}\hat{f}(x)\hat{\phi}(x)x^{\frac {\nu-2} {2}}\;dx
\end{equation}
The corresponding ultradistribution of exponential type in the one-dimensional  complex
variable $\rho$ is obtained in the following way: let $\hat{g}(t)$ defined as:
\begin{equation}
\label{ep3.5}
\hat{g}(t)=\frac {1} {(2\pi)^{\nu}}\int\limits_{-\infty}^{\infty}f(\rho)e^{-i\rho t}\;d\rho
\end{equation}
Then:
\begin{equation}
\label{ep3.6}
F(\rho)=\Theta[\Im(\rho)]\int\limits_0^{\infty}\hat{g}(t)e^{i\rho t}\;dt-
\Theta[-\Im(\rho)]\int\limits_{-\infty}^0\hat{g}(t)e^{i\rho t}\;dt
\end{equation}
or if we use Dirac's formula
\begin{equation}
\label{ep3.7}
F(\rho)=\frac {1} {2\pi i}\int\limits_{-\infty}^{\infty}\frac {f(t)} {t-\rho}\;dt
\end{equation}

The inversion formula $(\nu=2n)$ for $F(\rho)$ is given by
\begin{equation}
\label{ep3.8}
\hat{f}(x)=\frac {\pi} {(2\pi)^{\frac {\nu+2} {2}} x^{\frac {\nu-2} {4}}}
\oint\limits_{\Gamma} F(\rho) {\rho}^{\frac {\nu-2} {4}} {\cal J}_{\frac {\nu-2} {2}}
(x^{1/2}{\rho}^{1/2})\;d\rho
\end{equation}
Note that the part of the integrand that multiplies to
$F(\rho)$ is an entire function of $\rho$ for $\nu=2n$. In this case
the first term of (\ref{ep4.13}) take the form:
\begin{equation}
\label{ep3.9}
\oint\limits_{\Gamma}F(\rho)\phi(\rho){\rho}^{\frac {\nu-2} {2}}\;d\rho=
(2\pi)^{\nu}\int\limits_0^{\infty}\hat{f}(x)\hat{\phi}(x)x^{\frac {\nu-2} {2}}\;dx
\end{equation}

We give now same examples of the use of Fourier transform.

\subsection*{Examples}

As a first example we calculate the Fourier transform of
\begin{equation}
\label{ep3.10}
2^{-\nu}\Theta[\Im(z_1)]\Theta[\Im(z_2)]\cdot\cdot\cdot\Theta[\Im(z_{\nu})]
\cosh\left(a\sqrt{z_1^2+z_2^2+\cdot\cdot\cdot+z_{\nu}^2}\right)
\end{equation}
where a is a complex number for $\nu=2n$. From (\ref{ep3.3})
\[F(\rho)=\frac {(2\pi)^{\frac {\nu-2} {2}}} {{\rho}^{\frac {\nu-2} {4}}}
\int\limits_0^{\infty}\cosh(ax^{\frac {1} {2}})x^{\frac {\nu-2} {4}}\left\{
\Theta[\Im(\rho)]e^{-\frac {i\pi\nu} {4}}{\cal{K}}_{\frac {\nu-2} {2}}
(-ix^{1/2}{\rho}^{1/2})-\right.\]
\begin{equation}
\label{ep3.11}
\left.\Theta[-\Im(\rho)]e^{\frac {i\pi\nu} {4}}{\cal{K}}_{\frac {\nu-2} {2}}
(ix^{1/2}{\rho}^{1/2})\;dx\right\}
\end{equation}
Now:
\[\int\limits_0^{\infty}e^{ax^{1/2}}x^{\frac {\nu-2} {4}}
{\cal{K}}_{\frac {\nu-2} {2}} (-ix^{1/2}{\rho}^{1/2})=
2\sqrt{\pi}\;e^{\frac {i\pi(\nu+2)} {4}}
\frac {\Gamma(\nu)} {\Gamma(\frac {\nu+3} {2})} \frac {{\rho}^{\frac {\nu-2} {4}}}
{({\rho}^{1/2}-ia)}\;\times\]
\[\boldsymbol{F}\left(\nu ,\frac {\nu-1} {2} ,\frac {\nu+3} {2} ,\frac {a-i{\rho}^{1/2}} {a+i{\rho}^{1/2}}
\right)\,\,\,\,\,\Im(\rho)>0\]
\[\int\limits_0^{\infty}e^{ax^{1/2}}x^{\frac {\nu-2} {4}}
{\cal{K}}_{\frac {\nu-2} {2}} (ix^{1/2}{\rho}^{1/2})=
2\sqrt{\pi}\;e^{-\frac {i\pi(\nu+2)} {4}}
\frac {\Gamma(\nu)} {\Gamma(\frac {\nu+3} {2})} \frac {{\rho}^{\frac {\nu-2} {4}}}
{({\rho}^{1/2}+ia)}\;\times\]
\begin{equation}
\label{ep3.12}
\boldsymbol{F}\left(\nu ,\frac {\nu-1} {2} ,\frac {\nu+3} {2} ,\frac {a+i{\rho}^{1/2}} {a-i{\rho}^{1/2}}
\right)\,\,\,\,\,\Im(\rho)<0
\end{equation}
To obtain (\ref{ep3.12}) we have used $\boldsymbol{6.621}, (3)$ of ref.\cite{tp10}
(Here $\boldsymbol{F}$ is the hypergeometric function).
Then we have:
\[F(\rho)=(4\pi)^{\frac {\nu-2} {2}}i\frac {\Gamma(\nu)} {\Gamma(\frac {\nu+3} {2} )}
\left\{\frac {1} {({\rho}^{1/2}-ia)}
\boldsymbol{F}\left(\nu ,\frac {\nu-1} {2} ,\frac {\nu+3} {2} ,\frac {a-i{\rho}^{1/2}} {a+i{\rho}^{1/2}}
\right)\;+\right.\]
\begin{equation}
\label{ep3.13}
\left.\frac {1} {({\rho}^{1/2}+ia)}
\boldsymbol{F}\left(\nu ,\frac {\nu-1} {2} ,\frac {\nu+3} {2} ,\frac {a+i{\rho}^{1/2}} {a-i{\rho}^{1/2}}
\right)\right\}
\end{equation}
As a second example we evaluate the Fourier transform of
\[2^{-\nu}\Theta[\Im(z_1)]\Theta[\Im(z_2)]\cdot\cdot\cdot\Theta[\Im(z_{\nu})]
\frac {\pi\mu^{\frac {\nu -2} {2}}} {(2\pi)^{\frac {\nu +2} {2}}}
(z_1^2+z_2^2+\cdot\cdot\cdot+z_{\nu}^2)^{\frac {2-\nu} {2}} \times\]
\begin{equation}
\label{ep3.14}
{\cal{J}}_{\frac {\nu -2} {2}}[\mu(z_1^2+z_2^2+\cdot\cdot\cdot+z_{\nu}^2)^{\frac {1} {2}}]
\end{equation}
We take into account
that for $\nu$ even
${\cal{J}}_{\frac {\nu-2} {2}}=e^{\frac {i\pi(\nu-2)} {2}}
{\cal{J}}_{\frac {2-\nu} {2}}$. Thus:
\[F(\rho)=\frac {{\mu}^{\frac {\nu-2} {2}}} {4\pi}e^{\frac {i\pi(\nu-2)} {2}}
{\rho}^{\frac {2-\nu} {4}}\int\limits_0^{\infty}
{\cal{J}}_{\frac {2-\nu} {2}}(\mu x^{1/2})
\left\{\Theta[\Im(\rho)]e^{-\frac {i\pi\nu} {4}}
{\cal{K}}_{\frac {\nu-2} {2}}(-ix^{1/2}{\rho}^{1/2})\;-\right.\]
\begin{equation}
\label{ep3.15}
\left.\Theta[-\Im(\rho)]e^{\frac {i\pi\nu} {4}}
{\cal{K}}_{\frac {\nu-2} {2}}(ix^{1/2}{\rho}^{1/2})\right\}\;dx
\end{equation}
Now:
\[\int\limits_0^{\infty}{\cal{J}}_{\frac {2-\nu} {2}}(\mu x^{1/2})
{\cal{K}}_{\frac {\nu-2} {2}}(-ix^{1/2}{\rho}^{1/2})\;dx=
e^{\frac {i\pi(6-\nu)} {4}}{\mu}^{\frac {2-\nu} {2}}
\frac {{\rho}^{\frac {\nu-2} {4}}} {\rho-{\mu}^2}\;;\;\Im(\rho)>0\]
\begin{equation}
\label{ep3.16}
\int\limits_0^{\infty}{\cal{J}}_{\frac {2-\nu} {2}}(\mu x^{1/2})
{\cal{K}}_{\frac {\nu-2} {2}}(ix^{1/2}{\rho}^{1/2})\;dx=
e^{-\frac {i\pi(6-\nu)} {4}}{\mu}^{\frac {2-\nu} {2}}
\frac {{\rho}^{\frac {\nu-2} {4}}} {\rho-{\mu}^2}\;;\;\Im(\rho)<0
\end{equation}
where we have used $\boldsymbol{6.576}, (3)$ of ref.\cite{tp10}. Then we have:
\begin{equation}
\label{ep3.17}
F(\rho)=-\frac {1} {2\pi i (\rho-{\mu}^2)}
\end{equation}

\section{The Fourier Transform in Minkowskian Space}

\setcounter{equation}{0}

We define a Lorentz invariant ultradistribution of exponential type $\hat{F}(z)$
as an ultradistribution of exponential type such that $\hat{f}(t)$ in (\ref{ep2.23}) is
Lorentz invariant (Note that a Lorentz invariant
ultradistribution is not  necessarily Lorentz invariant in an explicit way).
In this case we can use for the Fourier transform of $\hat{f}(t)$
the formula obtained in ref.\cite{tp5}:

\[F(\rho)=(2\pi)^{\frac {\nu-2} {2}}\int\limits_{-\infty}^{\infty}\hat{f}(x)\left\{
\Theta[\Im(\rho)]
e^{\frac {i\pi(\nu-2)} {4}} \frac {(x+i0)^{\frac {\nu-2} {4}}} {{\rho}^{\frac {\nu-2} {4}}}
{\cal{K}}_{\frac {\nu-2} {2}}[-i(x+i0)^{1/2}{\rho}^{1/2}]\right.-\]
\begin{equation}
\label{ep4.1}
\left.\Theta[-\Im(\rho)]
e^{\frac {i\pi(2-\nu)} {4}} \frac {(x-i0)^{\frac {\nu-2} {4}}} {{\rho}^{\frac {\nu-2} {4}}}
{\cal{K}}_{\frac {\nu-2} {2}}[i(x-i0)^{1/2}{\rho}^{1/2}]\right\}\;dx
\end{equation}
Here we have taken $\rho=\gamma+i\sigma$ and
\begin{equation}
\label{ep4.2}
{\rho}^{1/2}=\sqrt{\frac {\gamma+\sqrt{{\gamma}^2+{\sigma}^2}} {2}}+iSgn(\sigma)
\sqrt{\frac {-\gamma+\sqrt{{\gamma}^2+{\sigma}^2}} {2}}
\end{equation}
When $\hat{f}$ is  Lorentz invariant , we can use
(\ref{ep4.1}) or adopt the following treatment: starting from
\begin{equation}
\label{ep4.3}
\iiiint\limits_{-\infty}^{\;\;\;\infty}f(\rho)\phi(\rho,k^0)\;d^4k=
(2\pi)^{\nu}\iiiint\limits_{-\infty}^{\;\;\;\infty}\hat{f}(x)\hat{\phi}(x,x^0)d^4x
\end{equation}
we can deduce the equality:
\[\iint\limits_{-\infty}^{\;\;\;\;\infty}f(\rho)\phi(\rho,k^0)
(k_0^2-\rho)_+^{\frac {\nu-3} {2}}\;d\rho\;dk^0=\]
\begin{equation}
\label{ep4.4}
\iint\limits_{-\infty}^{\;\;\;\;\infty}\hat{f}(x)\hat{\phi}(x,x^0)
(x-x_0^2)_+^{\frac {\nu-3} {2}}\;dx\;dx^0
\end{equation}
Let $g(t)$ be defined as:
\begin{equation}
\label{ep4.5}
\hat{g}(t)=\frac {1} {(2\pi)^{\nu}}\int\limits_{-\infty}^{\infty}f(\rho)e^{-i\rho t}\;d\rho
\end{equation}
Then:
\begin{equation}
\label{ep4.6}
F(\rho)=\Theta[\Im(\rho)]\int\limits_0^{\infty}\hat{g}(t)e^{i\rho t}\;dt-
\Theta[-\Im(\rho)]\int\limits_{-\infty}^0\hat{g}(t)e^{i\rho t}\;dt
\end{equation}
or if we use Dirac's formula
\begin{equation}
\label{ep4.7}
F(\rho)=\frac {1} {2\pi i}\int\limits_{-\infty}^{\infty}\frac {f(t)} {t-\rho}\;dt
\end{equation}
The inverse of the Fourier transform can be evaluated in the following way:
we define
\[\hat{G}(x,\Lambda)=\frac {1} {(2\pi)^{\frac {\nu+2} {2}}}\oint\limits_{\Gamma}F(\rho)\left\{
e^{\frac {i\pi(\nu-2)} {4}} \frac {(\rho+\Lambda)^{\frac {\nu-2} {4}}} {(x+i0)^{\frac {\nu-2} {4}}}
{\cal{K}}_{\frac {\nu-2} {2}}[-i(x+i0)^{1/2}(\rho+\Lambda)^{1/2}]\right.+\]
\begin{equation}
\label{ep4.8}
+\left.e^{\frac {i\pi(2-\nu)} {4}} \frac {(\rho-\Lambda)^{\frac {\nu-2} {4}}} {(x-i0)^{\frac {\nu-2} {4}}}
{\cal{K}}_{\frac {\nu-2} {2}}[i(x-i0)^{1/2}(\rho-\Lambda)^{1/2}]\right\}\;d\rho
\end{equation}
then
\begin{equation}
\label{ep4.9}
\hat{f}(x)=\hat{G}(x,i0^+)
\end{equation}
We give now same examples of the use of Fourier transform in
Minkowskian space.

\subsection*{Examples}

As a first example we consider the Fourier transform of the ultradistribution
\[2^{\nu}\Theta;\Im(z_0)]\Theta[\Im(z_2)]\cdot\cdot\cdot\Theta[\Im(z_{\nu-1})]
\left[\cosh\left(a\sqrt{|z_0^2-z_1^2-\cdot\cdot\cdot-z_{\nu-1}^2|}\right)\right.\]
\begin{equation}
\label{ep4.10}
+ \left.\cos\left(a\sqrt{|z_0^2-z_1^2-\cdot\cdot\cdot-z_{\nu-1}^2|}\right)\right]
\end{equation}
where a is a complex number.

The cut along the real axis of (\ref{ep4.10}) is:
\begin{equation}
\label{ep4.11}
2^{-1}\left[e^{a\sqrt{|x_0^2-r^2|}}+e^{-a\sqrt{|x_0^2-r^2|}}+
e^{ia\sqrt{|x_0^2-r^2|}}+e^{-ia\sqrt{|x_0^2-r^2|}}\right]
\end{equation}
The Fourier transform is:
\[F(\rho)=(2\pi)^{\frac {\nu-2} {2}}\int\limits_{-\infty}^{\infty}
e^{|x|^{\frac {1} {2}}}
2^{-1}\left[e^{a\sqrt{|x_0^2-r^2|}}+e^{-a\sqrt{|x_0^2-r^2|}}+
e^{ia\sqrt{|x_0^2-r^2|}}+e^{-ia\sqrt{|x_0^2-r^2|}}\right]\]
\[\left\{
\Theta[\Im(\rho)]
e^{\frac {i\pi(\nu-2)} {4}} \frac {(x+i0)^{\frac {\nu-2} {4}}} {{\rho}^{\frac {\nu-2} {4}}}
{\cal{K}}_{\frac {\nu-2} {2}}[-i(x+i0)^{1/2}{\rho}^{1/2}]\right.-\]
\begin{equation}
\label{ep4.12}
\left.\Theta[-\Im(\rho)]
e^{\frac {i\pi(2-\nu)} {4}} \frac {(x-i0)^{\frac {\nu-2} {4}}} {{\rho}^{\frac {\nu-2} {4}}}
{\cal{K}}_{\frac {\nu-2} {2}}[i(x-i0)^{1/2}{\rho}^{1/2}]\right\}\;dx
\end{equation}
Now:
\[e^{\frac {i\pi(\nu-2)}{4}}\int\limits_{-\infty}^{\infty}e^{a|x|^{\frac {1} {2}}}
(x+i0)^{\frac {\nu-2} {4}}{\cal{K}}_{\frac {\nu-2} {2}}[-i(x+i0)^{1/2}{\rho}^{1/2}]=\]
\[2^{\frac {\nu} {2}}\sqrt{\pi}\frac {\Gamma(\nu)} {\Gamma(\frac {\nu+3} {2})}
\frac {e^{\frac {i\pi\nu} {2}}} {({\rho}^{1/2}-ia)^{\nu}}
\boldsymbol{F}\left(\nu,\frac {\nu-1} {2},\frac {\nu+3} {2},
\frac {a-i{\rho}^{1/2}} {a+i{\rho}^{1/2}}\right)-\]
\begin{equation}
\label{ep4.13}
2^{\frac {\nu} {2}}\sqrt{\pi}\frac {\Gamma(\nu)} {\Gamma(\frac {\nu+3} {2})}
\frac {e^{\frac {i\pi\nu} {2}}} {({\rho}^{1/2}+a)^{\nu}}
\boldsymbol{F}\left(\nu,\frac {\nu-1} {2},\frac {\nu+3} {2},
\frac {a+{\rho}^{1/2}} {a-{\rho}^{1/2}}\right)\;\;\Im(\rho)>0
\end{equation}
\[e^{\frac {i\pi(2-\nu)} {4}}\int\limits_{-\infty}^{\infty}e^{a|x|^{\frac {1} {2}}}
(x-i0)^{\frac {\nu-2} {4}}{\cal{K}}_{\frac {\nu-2} {2}}[i(x-i0)^{1/2}{\rho}^{1/2}]=\]
\[2^{\frac {\nu} {2}}\sqrt{\pi}\frac {\Gamma(\nu)} {\Gamma(\frac {\nu+3} {2})}
\frac {e^{-\frac {i\pi\nu} {2}}} {({\rho}^{1/2}+ia)^{\nu}}
\boldsymbol{F}\left(\nu,\frac {\nu-1} {2},\frac {\nu+3} {2},
\frac {a+i{\rho}^{1/2}} {a-i{\rho}^{1/2}}\right)-\]
\begin{equation}
\label{ep4.14}
2^{\frac {\nu} {2}}\sqrt{\pi}\frac {\Gamma(\nu)} {\Gamma(\frac {\nu+3} {2})}
\frac {e^{\frac {i\pi\nu} {2}}} {({\rho}^{1/2}+a)^{\nu}}
\boldsymbol{F}\left(\nu,\frac {\nu-1} {2},\frac {\nu+3} {2},
\frac {a+{\rho}^{1/2}} {a-{\rho}^{1/2}}\right)\;\;\Im(\rho)<0
\end{equation}
For to obtain (\ref{ep4.13}) and (\ref{ep4.14}) we have used $\boldsymbol{6.621},(3)$
of ref.\cite{tp10}. With these results we have:
\[F(\rho)=
\frac {(4\pi)^{\frac {\nu-1} {2}}} {2}\frac {\Gamma(\nu)} {\Gamma(\frac {\nu+3} {2})}
\left\{\Theta[\Im(\rho)]e^{\frac {i\pi\nu} {2}}\left[
\frac {\boldsymbol{F}\left(\nu,\frac {\nu-1} {2},\frac {\nu+3} {2},
\frac {a-i{\rho}^{1/2}} {a+i{\rho}^{1/2}}\right)}
{({\rho}^{1/2}-ia)^{\nu}}\right.\right. +\]
\[\frac {\boldsymbol{F}\left(\nu,\frac {\nu-1} {2},\frac {\nu+3} {2},
\frac {a+i{\rho}^{1/2}} {a-i{\rho}^{1/2}}\right)}
{({\rho}^{1/2}+ia)^{\nu}}+
\frac {\boldsymbol{F}\left(\nu,\frac {\nu-1} {2},\frac {\nu+3} {2},
\frac {a-{\rho}^{1/2}} {a+{\rho}^{1/2}}\right)}
{({\rho}^{1/2}+a)^{\nu}}+
\frac {\boldsymbol{F}\left(\nu,\frac {\nu-1} {2},\frac {\nu+3} {2},
\frac {a+{\rho}^{1/2}} {a-{\rho}^{1/2}}\right)}
{({\rho}^{1/2}-a)^{\nu}}-\]
\[\frac {\boldsymbol{F}\left(\nu,\frac {\nu-1} {2},\frac {\nu+3} {2},
\frac {a+{\rho}^{1/2}} {a-{\rho}^{1/2}}\right)}
{({\rho}^{1/2}+a)^{\nu}}-
\frac {\boldsymbol{F}\left(\nu,\frac {\nu-1} {2},\frac {\nu+3} {2},
\frac {a-{\rho}^{1/2}} {a+{\rho}^{1/2}}\right)}
{({\rho}^{1/2}-a)^{\nu}}-
\frac {\boldsymbol{F}\left(\nu,\frac {\nu-1} {2},\frac {\nu+3} {2},
\frac {a-i{\rho}^{1/2}} {a+i{\rho}^{1/2}}\right)}
{({\rho}^{1/2}+ia)^{\nu}}-\]
\[\left.\frac {\boldsymbol{F}\left(\nu,\frac {\nu-1} {2},\frac {\nu+3} {2},
\frac {a+i{\rho}^{1/2}} {a-i{\rho}^{1/2}}\right)}
{({\rho}^{1/2}-ia)^{\nu}}\right]-\Theta[-\Im(\rho)]
e^{-\frac {i\pi\nu} {2}}\left[
\frac {\boldsymbol{F}\left(\nu,\frac {\nu-1} {2},\frac {\nu+3} {2},
\frac {a+i{\rho}^{1/2}} {a-i{\rho}^{1/2}}\right)}
{({\rho}^{1/2}+ia)^{\nu}}\right.+\]
\[\frac {\boldsymbol{F}\left(\nu,\frac {\nu-1} {2},\frac {\nu+3} {2},
\frac {a-i{\rho}^{1/2}} {a+i{\rho}^{1/2}}\right)}
{({\rho}^{1/2}-ia)^{\nu}}+
\frac {\boldsymbol{F}\left(\nu,\frac {\nu-1} {2},\frac {\nu+3} {2},
\frac {a+{\rho}^{1/2}} {a-{\rho}^{1/2}}\right)}
{({\rho}^{1/2}-a)^{\nu}}+
\frac {\boldsymbol{F}\left(\nu,\frac {\nu-1} {2},\frac {\nu+3} {2},
\frac {a-{\rho}^{1/2}} {a+{\rho}^{1/2}}\right)}
{({\rho}^{1/2}+a)^{\nu}}-\]
\[\frac {\boldsymbol{F}\left(\nu,\frac {\nu-1} {2},\frac {\nu+3} {2},
\frac {a+{\rho}^{1/2}} {a-{\rho}^{1/2}}\right)}
{({\rho}^{1/2}+a)^{\nu}}-
\frac {\boldsymbol{F}\left(\nu,\frac {\nu-1} {2},\frac {\nu+3} {2},
\frac {a-{\rho}^{1/2}} {a+{\rho}^{1/2}}\right)}
{({\rho}^{1/2}-a)^{\nu}}-
\frac {\boldsymbol{F}\left(\nu,\frac {\nu-1} {2},\frac {\nu+3} {2},
\frac {a-i{\rho}^{1/2}} {a+i{\rho}^{1/2}}\right)}
{({\rho}^{1/2}+ia)^{\nu}}-\]
\begin{equation}
\label{ep4.15}
\left.\left.
\frac {\boldsymbol{F}\left(\nu,\frac {\nu-1} {2},\frac {\nu+3} {2},
\frac {a+i{\rho}^{1/2}} {a-i{\rho}^{1/2}}\right)}
{({\rho}^{1/2}-ia)^{\nu}}\right]\right\}
\end{equation}
As a second example we evaluate the Fourier transform of the ultradistribution ($\nu=2n$):
\[\hat{F}(z)=-\frac {(-1)^{\frac {\nu} {2}}i{\mu}^{\nu-2}} {2^{\nu}{\pi}^{\frac {\nu-2} {2}}}
\Theta[\Im(z_0)]\Theta[\Im(z_2)]\cdot\cdot\cdot-\Theta[\Im(z_{\nu-1})]\times\]
\begin{equation}
\label{ep4.16}
\sum\limits_{k=0}^{\infty}
\frac {(-1)^k{\mu}^{2k}(z_0^2-z_1^2-\cdot\cdot\cdot-z_{\nu-1}^2)^{kk}}
{2^{2k}(k)!\Gamma(\nu+k)}
\end{equation}
The cut along the real axis of $\hat{F}(z)$ is:
\begin{equation}
\label{ep4.17}
\hat{f}(x)={\hat{f}}_{\mu}(x_+)-{\hat{f}}_{\mu}(x_-)
\end{equation}
where
\begin{equation}
\label{ep4.18}
{\hat{f}}_{\mu}(x)=-\frac {i\pi} {2} \frac {{\mu}^{\frac {\nu-2} {2}}}
{(2\pi)^{\frac {\nu} {2}}}x^{\frac {2-\nu} {4}} {\cal{J}}_{\frac {2-\nu} {2}}
(\mu x^{1/2})
\end{equation}
Observe that
\begin{equation}
\label{ep4.19}
{\hat{f}}_{\mu}(x_+)=w_{\mu}(x)=-\frac {i\pi} {2} \frac {{\mu}^{\frac {\nu-2} {2}}}
{(2\pi)^{\frac {\nu} {2}}}x_+^{\frac {2-\nu} {4}} {\cal{J}}_{\frac {2-\nu} {2}}
(\mu x_+^{1/2})
\end{equation}
is the complex mass Wheeler's propagator.
Then according to (\ref{ep4.1})
\[F(\rho)=-\frac {i(\mu)^{\frac {\nu-2} {2}}} {4}\int\limits_0^{\infty}
{\cal{J}}_{\frac {2-\nu} {2}}(\mu x^{1/2})\left\{
\frac {\Theta[\Im(\rho)]} {{\rho}^{\frac {\nu-2} {4}}}\left[
e^{\frac {i\pi(\nu-2)} {4}}
{\cal{K}}_{\frac {\nu-2} {2}}(-ix^{1/2}{\rho}^{1/2})\right.\right.+\]
\[\left. e^{\frac {i\pi(\nu-2)} {2}}
{\cal{K}}_{\frac {\nu-2} {2}}(x^{1/2}{\rho}^{1/2})\right]-
\Theta[-\Im(\rho)]\left[
e^{\frac {i\pi(2-\nu)} {4}}
{\cal{K}}_{\frac {\nu-2} {2}}(ix^{1/2}{\rho}^{1/2})\right.+\]
\begin{equation}
\label{ep4.20}
\left.\left. e^{\frac {i\pi(2-\nu)} {2}}
{\cal{K}}_{\frac {\nu-2} {2}}(x^{1/2}{\rho}^{1/2})\right]\right\}\;dx
\end{equation}
Taking into account that
(See $\boldsymbol{6.576},(3)$, ref.\cite{tp10}):
\[\int\limits_0^{\infty}
{\cal{J}}_{\frac {2-\nu} {2}}(x^{1/2})
{\cal{K}}_{\frac {\nu-2} {2}}(-ix^{1/2}{\rho}^{1/2})\;dx=2{\mu}^{\frac {2-\nu} {2}}
e^{\frac {i\pi(6-\nu)} {4}}\frac {{\rho}^{\frac {\nu-2} {4}}} {\rho-{\mu}^2}\;\Im(\rho)>0\]
\[\int\limits_0^{\infty}
{\cal{J}}_{\frac {2-\nu} {2}}(x^{1/2})
{\cal{K}}_{\frac {\nu-2} {2}}(ix^{1/2}{\rho}^{1/2})\;dx=2{\mu}^{\frac {2-\nu} {2}}
e^{\frac {i\pi(\nu-6)} {4}}\frac {{\rho}^{\frac {\nu-2} {4}}} {\rho-{\mu}^2}\;\Im(\rho)<0\]
\begin{equation}
\label{ep4.21}
\int\limits_0^{\infty}
{\cal{J}}_{\frac {2-\nu} {2}}(x^{1/2})
{\cal{K}}_{\frac {\nu-2} {2}}(x^{1/2}{\rho}^{1/2})\;dx=2{\mu}^{\frac {2-\nu} {2}}
\frac {{\rho}^{\frac {\nu-2} {4}}} {\rho+{\mu}^2}
\end{equation}
we obtain:
\begin{equation}
\label{ep4.22}
F(\rho)=\frac {i} {2} Sgn[\Im(\rho)] \left[\frac {1} {\rho-{\mu}^2}+
\frac {\cosh\pi(\frac {\nu-2} {2})} {\rho+{\mu}^2}\right]
\end{equation}

\section{The generalization of Dimensional regularization to Ultradistributions
of Exponential Type}

\setcounter{equation}{0}

Let $\hat{F}(z)$ and $\hat{G}(z)$ be ultradistributions of exponential type such that
their cuts  along  the real axis are $\hat{f}(x)$ and $\hat{g}(x)$.
We suppose that $\hat{F}(z)$ and $\hat{G}(z)$ are spherically symmetric in Euclidean case
or Lorentz Invariant in Minkowskian space.
If we use the dimension $\nu$ as a regularizing parameter we can define
the convolution of $F(\rho)$ and $G(\rho)$ as:
\begin{equation}
\label{ep5.1}
F(\rho,\nu)\ast G(\rho,\nu)= (2\pi)^{\nu} {\cal F}\left\{\hat{f}(x,\nu)\hat{g}(x,\nu)\right\}
\end{equation}

\subsection*{The Euclidean Case}

As an example of use of
 (\ref{ep5.1}) in Euclidean space we consider an ultradistribution of exponential type
$\hat{F}(z)$ such that $\hat{f}(x)$ is defined in the point $a>0$ of the real axis and take the value
$\hat{f}(a)$, with the ultradistribution $\hat{G}(z)$ whose cut along the real axis is $\delta(x-a)$
 According to (\ref{ep3.1}) we have
\begin{equation}
\label{ep5.2}
{\cal F}\left\{\hat{F}\right\}(\rho)=F(\rho)
\end{equation}
\[{\cal F}\left\{\delta(x-a)\right\}=G(\rho)=\frac {(2\pi)^{\frac {\nu-2} {2}}} {{\rho}^{\frac {\nu-2} {4}}}
a^{\frac {\nu-2} {4}}\left\{\Theta[\Im(\rho)]
e^{-\frac {i\pi\nu} {4}}{\cal{K}}_{\frac {\nu-2} {2}}(-ia^{1/2}{\rho}^{1/2})-\right.\]
\[\left.\Theta[-\Im(\rho)]
e^{\frac {i\pi\nu} {4}}{\cal{K}}_{\frac {\nu-2} {2}}(ia^{1/2}{\rho}^{1/2})\right\}\;\; +\]
\begin{equation}
\label{ep5.3}
\frac {2{\pi}^{\frac {\nu-2} {2}}} {\Gamma(\frac {\nu-2} {2})
{\rho}^{\frac {\nu-2} {4}}}
a^{\frac {\nu-2} {4}} {\cal{S}}_{\frac {\nu-4} {2},
\frac {\nu-2} {2}}(a^{1/2}{\rho}^{1/2})
\end{equation}
Due to
\begin{equation}
\label{ep5.4}
{\cal F}\left\{\hat{f}(x)\delta(x-a)\right\}=\hat{f}(a){\cal F}\left\{\delta(x-a)\right\}=
\hat{f}(a)G(\rho)
\end{equation}
we have
\begin{equation}
\label{ep5.5}
F(\rho)\ast G(\rho)=(2\pi)^{\nu}\hat{f}(a) G(\rho)
\end{equation}

\subsection*{The Minkowskian case}

We consider
\begin{equation}
\label{ep5.6}
F(\rho)=G(\rho)=\frac {i} {2} Sgn[\Im(\rho)] \left[\frac {1} {\rho-{\mu}^2}+
\frac {\cosh\pi(\frac {\nu-2} {2})} {\rho+{\mu}^2}\right]
\end{equation}
From (\ref{ep4.17}) we have
\begin{equation}
\label{ep5.7}
\hat{f}(x)=\hat{g}(x)=-\frac {i\pi} {2} \frac {{(-\mu)}^{\frac {\nu-2} {2}}}
{(2\pi)^{\frac {\nu} {2}}}\left[x_+^{\frac {2-\nu} {4}} {\cal{J}}_{\frac {\nu-2} {2}}
(\mu x_+^{1/2}) -
x_-^{\frac {2-\nu} {4}} {\cal{J}}_{\frac {\nu-2} {2}}
(\mu x_-^{1/2})\right]
\end{equation}
Then
\[F(\rho)\ast G(\rho)=\frac {(2\pi)^{\frac {\nu +1} {2}}} {2^{\frac {3\nu -1} {2}}}
\Gamma\left(\frac {3-\nu} {2}\right)e^{i\pi (\frac {\nu-2} {2})}
{\rho}^{\frac {2-\nu} {2}} Sgn[\Im(\rho)]\times\]
\begin{equation}
\label{ep5.8}
\left[ ({\rho}^2-2\rho {\mu}^2)^{\frac {\nu-3} {2}}+ ({\rho}^2+2\rho {\mu}^2)^{\frac {\nu-3} {2}}\right]
\end{equation}
To obtain (\ref{ep5.8}) we use
\[\int\limits_0^{\infty}
{\cal J}_{\frac {2-\nu} {2}}({\mu}_1x){\cal J}_{\frac {2-\nu} {2}}({\mu}_2x)
{\cal K}_{\frac {\nu-2} {2}}(xz)\;dx=\]
\begin{equation}
\label{ep5.9}
-\frac {1} {\sqrt{\pi}}
\frac {\Gamma\left(\frac {3-\nu} {2}\right)} {2^{\frac {3\nu-6} {2}}}
\frac {z^{\frac {2-\nu} {2}}} {({\mu}_1{\mu}_2)^{\frac {\nu-2} {2}}}
\left[(z^2+{\mu}_1^2+{\mu}_2^2)^2-4{\mu}_1^2{\mu}_2^2\right]^{\frac {
\nu-3} {2}}
\end{equation}
and to deduce (\ref{ep5.9}) we have used:
\begin{equation}
\label{ep5.10}
{\cal K}_{\frac {\nu-2} {2}}(xz)=\frac {1} {2}
\left(\frac {zx} {2}\right)^{\frac {\nu-2} {2}}
\int\limits_0^{\infty}t^{-\frac {\nu} {2}}
e^{-t-{\frac {z^2x^2} {4t}}}dt
\end{equation}
(See $\boldsymbol{8.432} (6)$ of ref.\cite{tp10}).

We proceed now to the calculation of the convolution of two
ultradistributions of exponential type.

\section{The Convolution of two Ultradistributions of Exponential Type}

\setcounter{equation}{0}

The convolution of two ultraditributions of exponential type can be defined with a
change in the formula obtained in ref.(\cite{tp4}) for tempered oltradistributions  Let
here be

\begin{equation}
\label{ep6.1}
H_{\gamma\lambda}(k)=\frac {i} {2\pi}
\oint\limits_{{\Gamma}_1}
\oint\limits_{{\Gamma}_2}
\frac {[2\cosh(\gamma k_1)]^{-\lambda} F(k_1)
[2\cosh(\gamma k_2)]^{-\lambda} G(k_2)}
{k-k_1-k_2}dk_1\;dk_2
\end{equation}
\[\mid \Im(k)\mid>\mid \Im(k_1)\mid+\mid\Im(k_2\mid\;;\;
\gamma <\min\left(\frac {\pi} {2\mid\Im(k_1)\mid};\frac {\pi} {2\mid\Im(k_2)\mid}\right)\]
With this value of $\gamma$ the hyperbolic cosines has not singularities
in the integration zone.
Again we have the Laurent ( or Taylor )
expansion:
\begin{equation}
\label{ep6.2}
H_{\gamma\lambda}(k)=\sum\limits_nH_{\gamma}^{(n)}(k){\lambda}^n
\end{equation}
where the sum might have terms with negative n. We now define
the convolution product as the $\lambda$-independent term of (\ref{ep6.2})
\begin{equation}
\label{ep6.3}
(F\ast G)(k)=H(k)=H_{\gamma}^{(0)}(k)=H^{(0)}(k)
\end{equation}
that is $\gamma$-independent

To see this we consider a typical integral term in (\ref{ep6.1})
\begin{equation}
\label{ep6.4}
I=\int\limits_c^{\infty}\frac {F(k+i\sigma)} {[\cosh\gamma(k+i\sigma)]^{\lambda}}\;dk
\end{equation}
with
\begin{equation}
\label{ep6.5}
\left| F(k)\right| \leq A\left| k\right|^pe^{p\left|\Re(k)\right|}
\end{equation}
Then I has the value
\[I=e^{i(p+1)}\left[\sum\limits_{n=1 ; n\neq (p-1,p+1)}^{\infty}
a_n(p,\sigma)\frac {e^{-c(n-p-1)}} {n-p-1}- a_{p-1}(p,\sigma)\frac {e^{2c}} {2}\right.\]
\begin{equation}
\label{ep6.6}
\left. +\frac {a_{p+1}(p,\sigma)} {\lambda\gamma}\right]
\end{equation}
Thus the $\lambda$-independent term of I does not depend of $\gamma$.
As (\ref{ep6.1}) is composed by sums and products of integrals of the type
(\ref{ep6.4}) we conclude that (\ref{ep6.3}) is true.

\subsection*{Examples}

As a first example we consider the convolution of two exponentials.
Let be:
\begin{equation}
\label{ep6.7}
F(k)=Sgn[\Im(k)]\frac {e^{ak}} {2}\;\;\;;\;\;\;G(k)=Sgn[\Im(k)]\frac {e^{bk}} {2}
\end{equation}
(a and b complex). Then
\[H_{\gamma\lambda}(k)=\frac {i} {8\pi}
\oint\limits_{{\Gamma}_1}
\oint\limits_{{\Gamma}_2}
\frac {Sgn[\Im(k_1)]e^{ak_1}Sgn[\Im(k_2)]e^{bk_2}}
{[2\cosh(\gamma k_1)]^{\lambda}[2\cosh(\gamma k_2)]^{\lambda} (k-k_1-k_2)}dk_1\;dk_2=\]
\begin{equation}
\label{ep6.8}
\frac {1} {2\pi} \iint\limits_{-\infty}^{\;\;\;\;\infty}
\frac {e^{ak_1} e^{bk_2}}
{[2\cosh(\gamma k_1)]^{\lambda}[2\cosh(\gamma k_2)]^{\lambda} (k-k_1-k_2)}dk_1\;dk_2
\end{equation}
or
\[H_{\gamma\lambda}(k)=\frac {\Theta[\Im(k)]} {2\pi}
\int\limits_0^{\infty}\iint\limits_{-\infty}^{\;\;\;\;\infty}
\frac {e^{ak_1}e^{bk_2}e^{i(k-k_1-k_2)t}}
{(e^{\gamma k_1}+e^{-\gamma k_1})^{\lambda}
(e^{\gamma k_2}+e^{-\gamma k_2})^{\lambda}}\;dk_1\;dk_2\;dt\]
\begin{equation}
\label{ep6.9}
-\frac {\Theta[-\Im(k)]} {2\pi}
\int\limits_{-\infty}^0\iint\limits_{-\infty}^{\;\;\;\;\infty}
\frac {e^{ak_1}e^{bk_2}e^{i(k-k_1-k_2)t}}
{(e^{\gamma k_1}+e^{-\gamma k_1})^{\lambda}
(e^{\gamma k_2}+e^{-\gamma k_2})^{\lambda}}\;dk_1\;dk_2\;dt
\end{equation}
To evaluate (\ref{ep6.9}) we take into account that
\[\int\limits_{-\infty}^{\infty}\frac {e^{(a-it)k_1}} {(e^{\gamma k_1} + e^{-\gamma k_1})^{\lambda}}\;dk_1=
\frac {1} {2\gamma} \int\limits_0^{\infty} \frac {y^{\frac {\gamma \lambda + a -it} {2\gamma} -1}}
{{(1+Y)}^{\lambda}}\; dy = \]
\begin{equation}
\label{ep6.10}
\frac {1} {2\gamma} \frac {\Gamma\left(\frac {\gamma \lambda + a-it} {2\gamma}\right)
\Gamma\left(\frac {\gamma\lambda-a+it} {2\gamma}\right)} {\Gamma(\lambda)}
\end{equation}
Then
\[H_{\gamma\lambda}(k)=\frac {1} {8\pi{\gamma}^2{\Gamma}^2{(\lambda)}}
\left\{\Theta[\Im(k)]\int\limits_0^{\infty}
\Gamma\left(\frac {\gamma \lambda + a-it} {2\gamma}\right)
\Gamma\left(\frac {\gamma\lambda-a+it} {2\gamma}\right)\right.\times\]
\[\Gamma\left(\frac {\gamma \lambda + b-it} {2\gamma}\right)
\Gamma\left(\frac {\gamma\lambda-b+it} {2\gamma}\right)e^{ikt}\;dt\;\;\; -\]
\[\Theta[-\Im(k)]\int\limits_{-\infty}^0
\Gamma\left(\frac {\gamma \lambda + a-it} {2\gamma}\right)
\Gamma\left(\frac {\gamma\lambda-a+it} {2\gamma}\right) \times\]
\begin{equation}
\label{ep6.11}
\left.\Gamma\left(\frac {\gamma \lambda + b-it} {2\gamma}\right)
\Gamma\left(\frac {\gamma\lambda-b+it} {2\gamma}\right)e^{ikt}\;dt \right\}
\end{equation}
and using the equality (\cite{tp10}, {\bf 3.381}, 4)
\begin{equation}
\label{ep6.12}
\int\limits_0^{\infty} x^{\nu-1}e^{-\mu x}\;dx={\mu}^{-\nu}\Gamma(\nu)
\end{equation}
and performing the integral in the variable $t$, we have for (\ref{ep6.11})
\[H_{\gamma\lambda}(k)=-\frac {1} {8\pi i {\gamma}^2 {\Gamma}^2(\lambda)}
\iiiint\limits_{\;\;0}^{\;\;\;\;\infty}
s_1^{\frac {\gamma\lambda+a} {2\gamma}-1}e^{-s_1}
s_2^{\frac {\gamma\lambda-a} {2\gamma}-1}e^{-s_2}
s_3^{\frac {\gamma\lambda+b} {2\gamma}-1}e^{-s_3}
s_4^{\frac {\gamma\lambda-b} {2\gamma}-1}e^{-s_4}\times\]
\begin{equation}
\label{ep6.13}
\frac {1} {k+\frac {1} {2\gamma}\ln(\frac {s_2s_4} {s_1s_3})}\;ds_1\;ds_2\;ds_3\;ds_4
\end{equation}
As
\begin{equation}
\label{ep6.14}
\frac {1} {k+\frac {1} {2\gamma}\ln(\frac {s_2s_4} {s_1s_3})}=
\sum\limits_{n=0}^{\infty}\frac {(ik_I)^n} {n!}
\frac {{\partial}^n} {\partial k_R^n}
\delta\left[k_R+\frac {1} {2\gamma}\ln\left(
\frac {s_2s_4} {s_1s_3}\right)\right]
\end{equation}
where ($k=k_R+ik_I$) and
\begin{equation}
\label{ep6.15}
\delta\left[k_R+\frac {1} {2\gamma}\ln\left(
\frac {s_2s_4} {s_1s_3}\right)\right]=
\frac {s_1s_3} {s_2} e^{-2\gamma k_R}
\delta\left(s_4-\frac {s_1s_3} {s_2}e^{-2\gamma k_R}\right)
\end{equation}
we obtain
\[H_{\gamma\lambda}(k)=-
\sum\limits_{n=0}^{\infty}\frac {(ik_I)^n} {n!}
\frac {{\partial}^n} {\partial k_R^n}
\frac {e^{k_R(b-\gamma\lambda)}} {4\pi i\gamma{\Gamma}^2(\lambda)}
\iiiint\limits_{\;\;0}^{\;\;\;\;\infty}s_1^{\frac {2\gamma\lambda+a-b} {2\gamma}-1} e^{-s_1}\;\times\]
\begin{equation}
\label{ep6.16}
s_2^{\frac {b-a} {2\gamma}-1} e^{-s_2}
s_3^{\lambda-1} e^{-s_3}
e^{-\left(\frac {s_1 s_3} {s_2}e^{-2\gamma k_R}\right)}\;ds_1\;ds_2\;ds_3\;ds_4
\end{equation}
After to evaluate the four-fold integral $H_{\gamma\lambda}$ take the form:
\[H_{\gamma\lambda}(k)=-\frac {e^{k(b+\gamma\lambda)}} {4\pi i\gamma\Gamma(2\lambda)}
\Gamma\left(\frac {a-b+2\gamma\lambda} {2\gamma}\right)
\Gamma\left(\frac {b-a+2\gamma\lambda} {2\gamma}\right)\times\]
\begin{equation}
\label{ep6.17}
F\left(\lambda+\frac {b-a} {2\gamma}, \lambda,2\lambda; 1-e^{2\gamma k}\right)
\end{equation}
When $a\neq b$
\begin{equation}
\label{ep6.18}
\lim_{\lambda\rightarrow 0}H_{\gamma\lambda}(k)=0
\end{equation}
When $a=b$
\begin{equation}
\label{ep6.19}
\lim_{\lambda\rightarrow 0}H_{\gamma\lambda}(k)=\frac {ke^{ka}} {2\pi i}\equiv 0
\end{equation}
and then $H(k)$ is the null ultradistribution
Thus we have finally:
\[Sgn[\Im(k)]\frac {e^{ak}}{2}\ast Sgn[\Im(k)]\frac {e^{bk}} {2}=0\]
and Fourier antitransforming
\begin{equation}
\label{ep6.20}
\delta(z-a)\delta(z-b)=0
\end{equation}
As a second example we consider the convolution of two complex Dirac's deltas:
\begin{equation}
\label{ep6.21}
F(k)=-\frac {1} {2\pi i (k-a)}\;\;\;;\;\;\;G(k)=-\frac {1} {2\pi i (k-b)}
\end{equation}
We have
\[H_{\gamma\lambda}(k)=\frac {i} {2\pi}
\oint\limits_{{\Gamma}_1}
\oint\limits_{{\Gamma}_2}
\frac {i} {2\pi (k_1-a)} \frac {i} {2\pi (k_2-b)}\times\]
\begin{equation}
\label{ep6.22}
\frac {[2\cosh(\gamma k_1)]^{-\lambda}
[2\cosh(\gamma k_2)]^{-\lambda}}
{k-k_1-k_2}dk_1\;dk_2=
\end{equation}
\begin{equation}
\label{ep6.23}
\frac {1} {2\pi} \frac {[2\cosh(\gamma a)]^{-\lambda}
[2\cosh(\gamma b)]^{-\lambda}}
{k-a-b}
\end{equation}
and as a consequence
\begin{equation}
\label{ep6.24}
H(k)=-\frac {1} {2\pi i} \frac {1} {k-a-b}
\end{equation}
or
\begin{equation}
\label{ep6.25}
\delta(k-a)\ast \delta(k-b)=\delta(k-a-b)
\end{equation}
and in the configuration space:
\begin{equation}
\label{ep6.26}
\frac {Sgn[\Im(z)]} {2} e^{az}\;\; \frac {Sgn[\Im(z)]} {2} e^{bz}=\frac {Sgn[\Im(z)]} {2} e^{(a+b)z}
\end{equation}
Formula (\ref{ep6.1}) can be generelized to $\nu$ dimensions:
\[H_{\gamma\lambda}(k)=\frac {i^{\nu}} {(2\pi)^{\nu}}
\oint\limits_{{\Gamma}_1}
\oint\limits_{{\Gamma}_2}
\frac {\prod\limits_{j=1}^{\nu}[2\cosh(\gamma_j  k_{1j}]^{-\lambda_j}
[2\cosh(\gamma_j k_{2j})]^{-\lambda_j}}
{\prod\limits_{j=1}^{\nu} (k_j-k_{1j}-k_{2j})}\times \]
\begin{equation}
\label{ep6.27}
F(k_1) G(k_2)\;\;d^{\nu}k_1\;d^{\nu}k_2
\end{equation}
\[\gamma_j <\min\left(\frac {\pi} {2\mid\Im(k_{1j})\mid};\frac {\pi} {2\mid\Im(k_{2j})\mid}\right)\]
As in the one-dimensional case
\begin{equation}
\label{ep6.28}
H_{\gamma\lambda}(k)=\sum\limits_{n_1, n_2,..,n_{\nu}}
{\lambda}_1^{n_1} {\lambda}_2^{n_2}\cdot\cdot\cdot {\lambda}_{\nu}^{n_{\nu}}
H^{(n_1+n_2+\cdot\cdot\cdot +n_{\nu})}(k)
\end{equation}
and again
\begin{equation}
\label{ep6.30}
(F\ast G)(k)=H(k)=H^{(0)}(k)
\end{equation}

\section{Solution to the question of the normalization of Gamow States
in Quantum Mechanics}

\setcounter{equation}{0}

As an application of the results of section 6, we give in this section a solution
to the question of normalization of Gamow states in Quantum Mechanics.
If we have a Gamow state that depends on $l+m$ variables
$\phi(k_1,k_2,...,k_l;{\rho}_1,{\rho}_2,...{\rho}_m)$,
and we wish to calculate
\begin{equation}
\label{ep7.1}
I(k_1,k_2,..,k_l)=\int\limits_{-\infty}^{\infty}\cdot\cdot\cdot
\int\limits_{-\infty}^{\infty}
|\phi|^2(k_1,k_2,..,k_l;{\rho}_1,{\rho}_2,..,{\rho}_m)\;d{\rho}_1\;d{\rho}_2
\cdot\cdot d{\rho}_m
\end{equation}
we define
\[\Phi(k_1,k_2,...,k_l;z_1,z_2,...,z_m)=\frac {1} {(2\pi i)^m}
\int\limits_{-\infty}^{\infty}\cdot\cdot\cdot\int\limits_{-\infty}^{\infty}\;\;\;
\times\]
\begin{equation}
\label{ep7.2}
\frac {|\phi|^2(k_1,k_2,...,k_l;{\rho}_1,{\rho}_2,...,{\rho}_m)}
{({\rho}_1-z_1)({\rho}_2-z_2)\cdot\cdot\cdot({\rho}_m-z_m)}
\;d{\rho}_1\;d{\rho}_2\cdot\cdot\cdot d{\rho}_m
\end{equation}
and
\[H_{{\gamma}_1{\gamma}_2...{\gamma}_m
{\lambda}_1{\lambda}_2...{\lambda}_m}
(k_1,k_2,...,k_l)=\oint\limits_{{\Gamma}_1}\cdot\cdot\cdot
\oint\limits_{{\Gamma}_m}\;\;\;\times\]
\begin{equation}
\label{ep7.3}
\frac {\Phi(k_1,k_2,...,k_l;z_1,z_2,...,z_m)}
{[\cosh({\gamma}_1z_1)]^{{\lambda}_1}
[\cosh({\gamma}_2z_2)]^{{\lambda}_2}...
[\cosh({\gamma}_mz_m)]^{{\lambda}_m}}
\;dz_1\;dz_2\cdot\cdot\cdot dz_m
\end{equation}
We have again the Laurent's expansion:
\[H_{{\gamma}_1{\gamma}_2...{\gamma}_m
{\lambda}_1{\lambda}_2...{\lambda}_m}
(k_1,k_2,...,k_l)=
\sum\limits_{n_1,n_2,...n_m}\times\]
\begin{equation}
\label{ep7.4}
{\lambda}_1^{n_1}{\lambda}_2^{n_2}...{\lambda}_m^{n_m}
H_{{\gamma}_1{\gamma}_2...{\gamma}_m}^{(n_1+n_2+...+n_m)}
(k_1,k_2,...,k_l)
\end{equation}
and as a consequence of section 6 we define:
\[I(k_1,k_2,...,k_l)=H(k_1,k_2,...,k_l)=H^{(0)}(k_1,k_2,...,k_l)=\]
\begin{equation}
\label{ep7.5}
H^{(0)}_{{\gamma}_1{\gamma}_2...{\gamma}_m}
(k_1,k_2,...,k_l)
\end{equation}

As an example of appilcation of (\ref{ep7.1}-\ref{ep7.5})
we evaluate
\begin{equation}
\label{ep7.6}
I(k)=\int\limits_0^{\infty} {\phi}_0^2(k,r)\;dr=
\int\limits_{-\infty}^{\infty} \Theta(r){\phi}_0^2(k,r)\;dr
\end{equation}
where ${\phi}_0(k,r)$ is the $l=0$ function corresponding to
the Square-Well potential used in ref.\cite{tp11}:
\begin{equation}
\label{ep7.7}
{\phi}_0(k,r)=
\begin{cases}
\frac {\sin(qr)} {q}& \text{if $r<a$}\\
\frac {\sin(qa)} {q} e^{ik(a-r)}&  \text{if $r>a$}
\end{cases}
\end{equation}
Here $q$ is given by:
\begin{equation}
\label{ep7.8}
q^2=\frac {2m} {{\hbar}^2}[E-V(r)]=k^2-\frac {2m} {{\hbar}^2}V(r)
\end{equation}
and
\begin{equation}
\label{ep7.9}
V(r)=
\begin{cases}
0& \text{if $r>a$}\\
-V_0& \text{if $r\geq a $}
\end{cases}
\end{equation}
We can write:
\begin{equation}
\label{ep7.10}
\phi(k,r)=\left[\Theta(r)-\Theta(r-a)\right]\frac {\sin(qr)} {q}+
\Theta(r-a)\frac {\sin(qa)} {q} e^{ik(a-r)}
\end{equation}
\begin{equation}
\label{ep7.11}
{\phi}^2(k,r)=\left[\Theta(r)-\Theta(r-a)\right]\frac {{\sin}^2(qr)} {q^2}+
\Theta(r-a)\frac {{\sin}^2(qa)} {q^2} e^{2ik(a-r)}
\end{equation}
and according to (\ref{ep7.2})
\[\Phi(k,z)=\frac {1} {2\pi i} \left[\ln(a-z)-\ln(z)\right]
\frac {{\sin}^2(qz)} {q^2}\;\;\;-\]
\begin{equation}
\label{ep7.12}
\frac {1} {2\pi i}\frac {{\sin}^2(qa)} {q^2}\ln(a-z)
e^{2ik(a-z)}
\end{equation}
Thus we have:
\[H_{\gamma\lambda}(k)=\frac {1} {2\pi i q^2}
\oint\limits_{\Gamma}
\frac {\ln(a-z)-\ln(z)} {[\cosh(\gamma z)]^{\lambda}}
{\sin}^2(qz)\;dz\;\;\;-\]
\[\frac {{\sin}^2(qa)} {2\pi i q^2}
e^{2ika}
\oint\limits_{\Gamma}
\frac {\ln(a-z)} {[\cosh(\gamma z)]^{\lambda}}
e^{-2ikz}\;dz\;=\]
\begin{equation}
\label{ep7.13}
\frac {1} {q^2}\int\limits_0^a
\frac {{\sin}^2(qr)} {[\cosh(\gamma r)]^{\lambda}}\;dr+
\frac {{\sin}^2(qa)} {q^2} e^{2ika}
\int\limits_a^{\infty}\frac {e^{-2ikr}}
{[\cosh(\gamma r)]^{\lambda}}\;dr
\end{equation}
We can evaluate the second integral in (\ref{ep7.13}):
\[\int\limits_a^{\infty}
\frac {e^{-2ikr}} {[\cosh(\gamma r)]^{\lambda}}\;dr\;=\]
\begin{equation}
\label{ep7.14}
\frac {e^{-a(\gamma\lambda+2ik)}}
{\gamma\lambda+2ik}
F\left(\lambda,\frac {\gamma\lambda+2ik} {2\gamma},
\frac {\gamma\lambda+2ik} {2\gamma}+1;-e^{4\gamma a}\right)
\end{equation}
Taking into account that:
\begin{equation}
\label{ep7.15}
\lim_{\lambda \rightarrow 0}
\frac {1} {\gamma\lambda+2ik}=
\begin{cases}
\frac {-i} {2(k-i0)}& \text{if $\Im(k)=0$}\\
\frac {-i} {2k}& \text{if $\Im(k)\neq 0$}
\end{cases}
\end{equation}
we obtain:
\[I(k)=H(k)=\frac {1} {q^2}\int\limits_0^a{\sin}^2(qr)\;dr+
\frac {{\sin}^2(qa)} {2ikq^2}=\]
\begin{equation}
\label{ep7.16}
\frac {a} {2q^2}-\frac {\sin(2qa)} {4q^3}+
\frac {{\sin}^2(qa)} {2ikq^2}
\end{equation}
Using the equality:
\begin{equation}
\label{ep7.17}
\cos(qa)=-i\frac {k} {q} \sin(qa)
\end{equation}
we have:
\begin{equation}
\label{ep7.18}
I(k)=\frac {1+ika} {2ik}\;\frac {q^2-k^2} {q^4}\;{\sin}^2(qa)
\end{equation}
This result coincides with the result obtained in
ref.\cite{tp11}.

\section{The Convolution of four-dimensional even
Ultradistributionsof Exponential Type}

\setcounter{equation}{0}

The convolution of two even ultraditributions of exponential type can be defined with a
change of the formula obtained in ref.(\cite{tp4}) for tempered ultradistributions  Let
here be
\[H_{\gamma_0\gamma\lambda_0\lambda}
(k^0,\rho)=\frac {1} {4\pi\rho}\oint\limits_{{\Gamma}^0_1}
\oint\limits_{{\Gamma}^0_2}\oint\limits_{{\Gamma}_1}
\oint\limits_{{\Gamma}_2}
\frac {[2 cosh({\gamma}_0 k_1^0)]^{-\lambda_0}[2 cosh({\gamma}{\rho}_1)]^{-\lambda}
 F(k^0_1,{\rho}_1)}
{k^0-k^0_1-k^0_2}\times\]
\[[2 cosh({\gamma}_0 k_2^0)]^{-\lambda_0}[2 cosh({\gamma}{\rho}_2)]^{-\lambda}G(k^0_2,{\rho}_2)
\;\;\times\]
\begin{equation}
\label{ep8.1}
\ln [{\rho}^2-({\rho}_1+{\rho}_2)^2]
{\rho}_1{\rho}_2\;d{\rho}_1\;d{\rho}_2\;dk^0_1\;dk^0_2
\end{equation}
\[\mid\Im(k^0)\mid>\mid\Im(k_1^0\mid+\mid\Im(k_2^0)\mid\;;\;
\mid\Im(\rho)\mid>\mid\Im(\rho_1\mid+\mid\Im(\rho_2)\mid\]
\[\gamma_0 <\min\left(\frac {\pi} {2\mid\Im(k_1^0)\mid};\frac {\pi} {2\mid\Im(k_2^0)\mid}\right)\;;\;
\gamma <\min\left(\frac {\pi} {2\mid\Im(\rho_1)\mid};\frac {\pi} {2\mid\Im(\rho_2)\mid}\right)\]
The difference between
\[\int\frac {2\rho} {{\rho}^2-({\rho}_1+{\rho}_2)^2}\;d\rho\;\;\;\;\;\;
\mathrm{and}\;\;\;\;\;\;\ln[{\rho}^2-({\rho}_1+{\rho}_2)^2]\]
is an entire analytic function. With this substitution in
(\ref{ep8.1}) we obtain
\[H_{\gamma_0\gamma\lambda_0\lambda}
(k^0,\rho)=\frac {1} {2\pi\rho}\int \rho\;d\rho\oint\limits_{{\Gamma}^0_1}
\oint\limits_{{\Gamma}^0_2}\oint\limits_{{\Gamma}_1}
\oint\limits_{{\Gamma}_2}
\frac {F(k^0_1,{\rho}_1) G(k^0_2,{\rho}_2)}
{k^0-k^0_1-k^0_2}\times\]
\[[2 cosh({\gamma}_0 k_1^0)]^{-\lambda_0}[2 cosh({\gamma} {\rho}_1)]^{-\lambda}
[2 cosh({\gamma}_0 k_2^0)]^{-\lambda_0}[2 cosh({\gamma} {\rho}_2)]^{-\lambda}
\times\]
\begin{equation}
\label{ep8.2}
\frac {1} {\rho^2-(\rho_1+\rho_2)^2}\;\;\;
{\rho}_1{\rho}_2\;d{\rho}_1\;d{\rho}_2\;dk^0_1\;dk^0_2
\end{equation}
We can again perform the Laurent expansion :
\begin{equation}
\label{ep8.3}
H_{\gamma_0\gamma{\lambda}_0\lambda}(k^0,\rho)=
\sum\limits_{mn} H_{\gamma_0\gamma}^{(m,n)}(k^0,\rho){\lambda}^{m}_0{\lambda}^n
\end{equation}
and define the convolution product as the $({\lambda}_0,\lambda)$-independent
term of (\ref{ep8.3}).
\begin{equation}
\label{ep8.4}
H(k)=H(k^0,\rho)=H_{\gamma_0\gamma}^{(0,0)}(k^0,\rho)=H^{(0,0)}(k^0,\rho)
\end{equation}

If we define:
\[L_{\gamma_0\gamma\lambda_0\lambda}
(k^0,\rho)=\oint\limits_{{\Gamma}^0_1}
\oint\limits_{{\Gamma}^0_2}\oint\limits_{{\Gamma}_1}
\oint\limits_{{\Gamma}_2}
\frac {F(k^0_1,{\rho}_1) G(k^0_2,{\rho}_2)}
{k^0-k^0_1-k^0_2}\times\]
\[[2 cosh({\gamma}_0 k_1^0)]^{-\lambda_0}[2 cosh({\gamma} {\rho}_1)]^{-\lambda}
[2 cosh({\gamma}_0 k_2^0)]^{-\lambda_0}[2 cosh({\gamma} {\rho}_2)]^{-\lambda}
\times\]
\begin{equation}
\label{ep8.5}
\frac {1} {\rho^2-(\rho_1+\rho_2)^2}\;\;\;
{\rho}_1{\rho}_2\;d{\rho}_1\;d{\rho}_2\;dk^0_1\;dk^0_2
\end{equation}
then
\begin{equation}
\label{ep8.6}
H_{\gamma_0\gamma{\lambda}_0\lambda}(k^0,\rho)=\frac {1} {2\pi\rho}\int
L_{\gamma_0\gamma{\lambda}_0\lambda}(k^0,\rho)\;\rho\;d\rho
\end{equation}
Now we show that the cut on the real axis of
(\ref{ep8.1}) $h_{{\lambda}_0\lambda}(k^0,\rho)$
 is an even function of $k^0$ and  $\rho$.
It is explicitly odd in $\rho$. For the variable $k^0$
we take into  account that $e^{i\pi{\lambda}_0\{Sgn[\Im(k^0_1)]+
Sgn[\Im(k^0_2)]\}}=1$ and as a consequence (\ref{ep8.1}) is odd
in $k^0$ too. We consider now the parity in variable $\rho$.
\[\oint\limits_{{\Gamma}_0}\oint\limits_{\Gamma}
H_{{\lambda}_0\lambda}(k^0,-\rho)\phi(k^0,\rho)\;dk^0\;d{\rho}=
-\iint\limits_{-\infty}^{+\infty}
h_{{\lambda}_0\lambda}(k^0,-\rho)\phi(k^0,\rho)\;dk^0\;d{\rho}=\]
\begin{equation}
\label{ep8.7}
-\oint\limits_{{\Gamma}_0}\oint\limits_{\Gamma}
H_{{\lambda}_0\lambda}(k^0,\rho)\phi(k^0,\rho)\;dk^0\;d{\rho}=
-\iint\limits_{-\infty}^{+\infty}
h_{{\lambda}_0\lambda}(k^0,\rho)\phi(k^0,\rho)\;dk^0\;d{\rho}
\end{equation}
Thus we have
\begin{equation}
\label{ep8.8}
h_{{\lambda}_0\lambda}(k^0,-\rho)=h_{{\lambda}_0\lambda}(k^0,\rho)
\end{equation}
The proof for the variable $k^0$ is similar.

\subsection*{Examples}

As a first example we shall calculate the convolution between
$F(k_0,\rho)=\delta(k_0^2-a^2)\delta(\rho-b)$ and
$G(k_0,\rho)=\delta(k_0^2-c^2)\delta(\rho-d)$.
We have:
\[H(k_0,\rho)=\frac {bd} {16\pi |a||b|\rho}\left(\frac {1} {k_0-a-c}+
 \frac {1} {k_0-a+c} + \frac {1} {k_0+a-c} + \frac {1} {k_0+a+c}\right)
\times\]
\begin{equation}
\label{ep8.9}
\ln[{\rho}^2-(b+d)^2]
\end{equation}
and simplifying the last expression:
\[H(k_0,\rho)=\frac {bd} {8\pi |a||b|\rho}\left[
\frac {k_0-a} {(k_0-a)^2-c^2}+\frac {k_0+a} {(k_0+a)^2-c^2}\right]\times\]
\begin{equation}
\label{ep8.10}
\ln[{\rho}^2-(b+d)^2]
\end{equation}
As a second example we evaluate the convolution of
$F(k_0,\rho)=\delta(k_0)\delta(\rho-a)$ and
$G(k_0,\rho)=\frac {1} {2}  Sgn[\Im(k_0)]
e^{ibk_0}\delta(\rho-c)$

We have:
\[H_{{\gamma}_0\gamma{\lambda}_0\lambda}(k_0,\rho)=
\frac {ac} {8\pi \rho} \frac {\ln[{\rho}^2-(a+c)^2]}
{[\cosh(\gamma a]^{\lambda}[\cosh(\gamma c)]^{\lambda}}
\oint\limits_{{\Gamma}_{02}}
\frac {Sgn[\Im(k_{02})]e^{ibk_{02}}}
{[\cosh({\gamma}_0k_{02})]^{{\lambda}_0}
(k_0-k_{02})}\;dk_{02}\]
\begin{equation}
\label{ep8.11}
=\frac {ac} {4\pi \rho} \frac {\ln[{\rho}^2-(a+c)^2]}
{[\cosh(\gamma a]^{\lambda}[\cosh(\gamma c)]^{\lambda}}
\int\limits_{-\infty}^{\infty}
\frac {e^{ibk_{02}}}
{[\cosh({\gamma}_0k_{02})]^{{\lambda}_0}
(k_0-k_{02})}\;dk_{02}
\end{equation}
Taking into account that
\[\lim {\lambda}_0\rightarrow 0
\int\limits_{-\infty}^{\infty}
\frac {e^{ibk_{02}}}
{[\cosh({\gamma}_0k_{02})]^{{\lambda}_0}
(k_0-k_{02})}\;dk_{02}=\]
\begin{equation}
\label{ep8.12}
-\pi i Sgn[\Im(k_0)]e^{ibk_0}
\end{equation}
we obtain
\begin{equation}
\label{ep8.13}
H(k_0,\rho)=\frac {ac} {4\pi i \rho} Sgn[\Im(k_0)]
e^{ibk_0} \ln[{\rho}^2-(a+d)^2]
\end{equation}

\section{The Convolution of Spherically Symmetric Ultradistributions
of Exponential Type in Euclidean Space}

\setcounter{equation}{0}

The convolution of two spherically symmetric
Ultraditributions of Exponential Type can be defined with a
change of the formula obtained in ref.(\cite{tp5}) for tempered ultradistributions  Let
here be
\[H_{\gamma\lambda}(\rho)=\frac {i\pi} {4\rho}\oint\limits_{{\Gamma}_1}
\oint\limits_{{\Gamma}_2}[2\cosh(\gamma\rho_1)]^{-\lambda}F({\rho}_1)
[2\cosh(\gamma\rho_2)]^{-\lambda}G({\rho}_2) \;\times\]
\begin{equation}
\label{ep9.1}
\left[\rho-{\rho}_1-{\rho}_2-
\sqrt{(\rho-{\rho}_1-{\rho}_2)^2-4{\rho}_1{\rho}_2}\right]d{\rho}_1\;d{\rho}_2
\end{equation}
Again we have the Laurent expansion:
\begin{equation}
\label{ep9.2}
H_{\gamma\lambda}(\rho)=\sum\limits_{n=-m}^{\infty}H_{\gamma}^{(n)}(\rho){\lambda}^n
\end{equation}
We now define
the convolution product as the $\lambda$-independent term of
(\ref{ep9.2}):
\begin{equation}
\label{ep9.3}
H(\rho)=H_{\gamma}^{(0)}(\rho)=H^{(0)}(\rho)
\end{equation}
Let ${\hat{H}}_{\gamma\lambda}(x)$ be the Fourier antitransform of $H_{\gamma\lambda}(\rho)$:
\begin{equation}
\label{ep9.4}
{\hat{H}}_{\gamma\lambda}(x)=\sum\limits_{n=-m}^{\infty}{\hat{H}}_{\gamma}^{(n)}(x){\lambda}^n
\end{equation}
If we define:
\[{\hat{f}}_{\gamma\lambda}(x)={\cal F}^{-1}\{[2\cosh(\gamma\rho)]^{-\lambda}F(\rho)\}\]
\begin{equation}
\label{ep9.5}
{\hat{g}}_{\gamma\lambda}(x)={\cal F}^{-1}\{[2\cosh(\gamma\rho)]^{-\lambda}G(\rho)\}
\end{equation}
then
\begin{equation}
\label{ep9.6}
{\hat{H}}_{\gamma\lambda}(x)=(2\pi)^4{\hat{f}}_{\gamma\lambda}(x){\hat{g}}_{\gamma\lambda}(x)
\end{equation}
and with the use of the Laurent's developments of $\hat{f}$ and
$\hat{g}$:
\[{\hat{f}}_{\gamma\lambda}(x)=\sum\limits_{n=-m_f}^{\infty}{\hat{f}}_{\gamma}^{(n)}(x){\lambda}^n\]
\begin{equation}
\label{ep9.7}
{\hat{g}}_{\gamma\lambda}(x)=\sum\limits_{n=-m_f}^{\infty}{\hat{g}}_{\gamma}^{(n)}(x){\lambda}^n
\end{equation}
we can write:
\begin{equation}
\label{ep9.8}
\sum\limits_{n=-m}^{\infty}{\hat{H}}_{\gamma}^{(n)}(x){\lambda}^n=(2\pi)^4
\sum\limits_{n=-m}^{\infty}\left(\sum\limits_{k=-m}^n {\hat{f}}_{\gamma}^{(k)}(x)
{\hat{g}}_{\gamma}^{(n-k)}(x)\right){\lambda}^n
\end{equation}
$(m=m_f+m_g)$\\
and as a consequence:
\begin{equation}
\label{ep9.9}
{\hat{H}}^{(0)}(x)=
\sum\limits_{k=-m}^0{\hat{f}}_{\gamma}^{(k)}(x){\hat{g}}_{\gamma}^{(n-k)}(x)
\end{equation}
We shall  give now some examples.

\subsection*{Examples}

The first example that we shall give is the convolution between
$F[\rho)=\delta(\rho -a)$ and $G(\rho)=\delta(\rho -b)$
We have:
\[H_{\gamma\lambda}(\rho)=\frac {i\pi} {4\rho}\oint\limits_{{\Gamma}_1}
\oint\limits_{{\Gamma}_2}[2\cosh(\gamma\rho_1)]^{-\lambda}\delta({\rho}_1-a)
[2\cosh(\gamma\rho_2)]^{-\lambda}\delta({\rho}_2-b) \;\times\]
\begin{equation}
\label{ep9.10}
\left[\rho-{\rho}_1-{\rho}_2-
\sqrt{(\rho-{\rho}_1-{\rho}_2)^2-4{\rho}_1{\rho}_2}\right]d{\rho}_1\;d{\rho}_2
\end{equation}
whose result is
\begin{equation}
\label{ep9.11}
H(\rho)=\frac {i\pi} {4\rho} \left[\rho-a-b-
\sqrt{(\rho-a-b)^2-4ab}\right]
\end{equation}
When a and b are real numbers, from (\ref{ep9.11})  we obtain
in the real $\rho$-axis
\begin{equation}
\label{ep9.12}
h(\rho)=\frac {\pi} {2\rho}
\left[(\rho-a-b)^2-4ab\right]_{+}^{\frac {1} {2}}
\end{equation}
As a second example we evaluate the convolution between
$F(\rho)=E_i(-ia\rho)e^{ia\rho}/2\pi i$ and
$G(\rho)={\delta}^{'}(\rho)=(2\pi i {\rho}^2)^{-1}$,
where $E_i(z)$ is the Exponential Integral Function.
We have
\[H_{\gamma\lambda}(\rho)=\frac {1} {8\rho}
\oint\limits_{{\Gamma}_1}\oint\limits_{{\Gamma}_2}
\frac {E_i(-ia{\rho}_1)e^{ia{\rho}_1}{\delta}^{'}({\rho}_2)}
{[2\cosh(\gamma{\rho}_1)]^{\lambda}
[2\cosh(\gamma{\rho}_2)]^{\lambda}}\]
\begin{equation}
\label{ep9.13}
\left[ \rho - {\rho}_1-{\rho}_2-
\sqrt {(\rho -{\rho}_1-{\rho}_1)^2-4{\rho}_1{\rho}_2}\right]
\;d{\rho}_1\;d{\rho}_2
\end{equation}
After integration in the variable ${\rho}_2$ we have
\[H_{\gamma\lambda}(\rho)=\frac {1} {4\rho}
\oint\limits_{{\Gamma}_1}
\frac {{\rho}_1 E_i(-ia{\rho}_1)e^{ia{\rho}_1}}
{[2\cosh(\gamma{\rho}_1)]^{\lambda}({\rho}_1-\rho)}\;d{\rho}_1=\]
\[\frac {1} {4\rho}\int\limits_0^{\infty}
\frac {{\rho}_1 e^{ia{\rho}_1}}
{[2\cosh(\gamma{\rho}_1)]^{\lambda}({\rho}_1-\rho)}\;d{\rho}_1=\]
\begin{equation}
\label{ep9.14}
\frac {1} {4\rho}\int\limits_0^{\infty}
\frac {e^{ia{\rho}_1}}
{[2\cosh(\gamma{\rho}_1)]^{\lambda}}\;d{\rho}_1+
\frac {1} {4}\int\limits_0^{\infty}
\frac {e^{ia{\rho}_1}}
{[2\cosh(\gamma{\rho}_1)]^{\lambda}({\rho}_1-\rho)}\;d{\rho}_1
\end{equation}
After to evaluate the integrals in (\ref{ep9.15}) ($\lambda\rightarrow 0$)
we obtain:
\begin{equation}
\label{ep9.15}
H(\rho)=\frac {i} {4a\rho} -\frac {i} {8\pi}e^{ia\rho}
E_i(-ia\rho)
\end{equation}

\section{The Convolution of two Lorentz invariant
Ultradistributions of Exponential Type in Minkowskian space}

\setcounter{equation}{0}

For Lorentz invariant ultradistributions of exponential type, following
ref.(\cite{tp5}) we have:
\[H_{\gamma\lambda}(\rho, \Lambda)=\frac {1} {8{\pi}^2\rho}
\int\limits_{{\Gamma}_1}\int\limits_{{\Gamma}_2}
[2\cosh(\gamma\rho_1)]^{-\lambda}F({\rho}_1)
[2\cosh(\gamma\rho_2)]^{-\lambda}G({\rho}_2)\]
\[\left\{\Theta[\Im(\rho)]\left\{[
\ln(-{\rho}_1+\Lambda)-\ln(-{\rho}_1-\Lambda)]\times
\right.\right.\]
\[[\ln(-{\rho}_2+\Lambda)-\ln(-{\rho}_2-\Lambda)]
\sqrt{4({\rho}_1+\Lambda)({\rho}_2+\Lambda)-
(\rho-{\rho}_1-{\rho}_2-2\Lambda)^2}\times\]
\[\ln\left[\frac {\sqrt{4({\rho}_1+\Lambda)({\rho}_2+\Lambda)-
(\rho-{\rho}_1-{\rho}_2-2\Lambda)^2}-i(\rho-{\rho}_1-{\rho}_2-2\Lambda)}
{2\sqrt{({\rho}_1+\Lambda)({\rho}_2+\Lambda)}}\right]+ \]
\[[\ln({\rho}_1+\Lambda)-\ln({\rho}_1-\Lambda)]
[\ln({\rho}_2+\Lambda)-\ln({\rho}_2-\Lambda)]\times\]
\[\sqrt{4({\rho}_1-\Lambda)({\rho}_2-\Lambda)-
(\rho-{\rho}_1-{\rho}_2+2\Lambda)^2}\times\]
\[\ln\left[\frac {\sqrt{4({\rho}_1-\Lambda)({\rho}_2-\Lambda)-
(\rho-{\rho}_1-{\rho}_2+2\Lambda)^2}-i(\rho-{\rho}_1-{\rho}_2+2\Lambda)}
{2\sqrt{({\rho}_1-\Lambda)({\rho}_2-\Lambda)}}\right]+ \]
\[[\ln({\rho}_1+\Lambda)-\ln({\rho}_1-\Lambda)]
[\ln(-{\rho}_2+\Lambda)-\ln(-{\rho}_2-\Lambda)]\times\]
\[\left\{\frac {i\pi} {2}\left[
\sqrt{4({\rho}_1+\Lambda)({\rho}_2-\Lambda)-
(\rho-{\rho}_1-{\rho}_2)^2}-i(\rho-{\rho}_1-{\rho}_2)\right]\right.+\]
\[\sqrt{4({\rho}_1+\Lambda)({\rho}_2-\Lambda)-
(\rho-{\rho}_1-{\rho}_2)^2}\times\]
\[\left.\ln\left[\frac {\sqrt{4({\rho}_1+\Lambda)({\rho}_2-\Lambda)-
(\rho-{\rho}_1-{\rho}_2)^2}-i(\rho-{\rho}_1-{\rho}_2)}
{2i\sqrt{-({\rho}_1+\Lambda)({\rho}_2-\Lambda)}}\right]\right\}+ \]
\[[\ln(-{\rho}_1+\Lambda)-\ln(-{\rho}_1-\Lambda)]
[\ln({\rho}_2+\Lambda)-\ln({\rho}_2-\Lambda)]\times\]
\[\left\{\frac {i\pi} {2}\left[
\sqrt{4({\rho}_1-\Lambda)({\rho}_2+\Lambda)-
(\rho-{\rho}_1-{\rho}_2)^2}-i(\rho-{\rho}_1-{\rho}_2)\right]\right.+\]
\[\sqrt{4({\rho}_1-\Lambda)({\rho}_2+\Lambda)-
(\rho-{\rho}_1-{\rho}_2)^2}\times\]
\[\left.\left.\ln\left[\frac {\sqrt{4({\rho}_1-\Lambda)({\rho}_2+\Lambda)-
(\rho-{\rho}_1-{\rho}_2)^2}-i(\rho-{\rho}_1-{\rho}_2)}
{2i\sqrt{-({\rho}_1-\Lambda)({\rho}_2+\Lambda)}}\right]\right\}\right\}- \]
\[\Theta[-\Im(\rho)]\left\{[
\ln(-{\rho}_1+\Lambda)-\ln(-{\rho}_1-\Lambda)]
[\ln(-{\rho}_2+\Lambda)-\ln(-{\rho}_2-\Lambda)]
\times\right.\]
\[\sqrt{4({\rho}_1-\Lambda)({\rho}_2-\Lambda)-
(\rho-{\rho}_1-{\rho}_2+2\Lambda)^2}\times\]
\[\ln\left[\frac {\sqrt{4({\rho}_1-\Lambda)({\rho}_2-\Lambda)-
(\rho-{\rho}_1-{\rho}_2+2\Lambda)^2}-i(\rho-{\rho}_1-{\rho}_2+2\Lambda)}
{2\sqrt{({\rho}_1-\Lambda)({\rho}_2-\Lambda)}}\right]+ \]
\[[\ln({\rho}_1+\Lambda)-\ln({\rho}_1-\Lambda)]
[\ln({\rho}_2+\Lambda)-\ln({\rho}_2-\Lambda)]\times\]
\[\sqrt{4({\rho}_1+\Lambda)({\rho}_2+\Lambda)-
(\rho-{\rho}_1-{\rho}_2-2\Lambda)^2}\times\]
\[\ln\left[\frac {\sqrt{4({\rho}_1+\Lambda)({\rho}_2+\Lambda)-
(\rho-{\rho}_1-{\rho}_2-2\Lambda)^2}-i(\rho-{\rho}_1-{\rho}_2-2\Lambda)}
{2\sqrt{({\rho}_1+\Lambda)({\rho}_2+\Lambda)}}\right]+ \]
\[[\ln({\rho}_1+\Lambda)-\ln({\rho}_1-\Lambda)]
[\ln(-{\rho}_2+\Lambda)-\ln(-{\rho}_2-\Lambda)]\times\]
\[\left\{\frac {i\pi} {2}\left[
\sqrt{4({\rho}_1-\Lambda)({\rho}_2+\Lambda)-
(\rho-{\rho}_1-{\rho}_2)^2}-i(\rho-{\rho}_1-{\rho}_2)\right]\right.+\]
\[\sqrt{4({\rho}_1-\Lambda)({\rho}_2+\Lambda)-
(\rho-{\rho}_1-{\rho}_2)^2}\times\]
\[\left.\ln\left[\frac {\sqrt{4({\rho}_1-\Lambda)({\rho}_2+\Lambda)-
(\rho-{\rho}_1-{\rho}_2)^2}-i(\rho-{\rho}_1-{\rho}_2)}
{2i\sqrt{-({\rho}_1-\Lambda)({\rho}_2+\Lambda)}}\right]\right\}+ \]
\[[\ln(-{\rho}_1+\Lambda)-\ln(-{\rho}_1-\Lambda)]
[\ln({\rho}_2+\Lambda)-\ln({\rho}_2-\Lambda)]\times\]
\[\left\{\frac {i\pi} {2}\left[
\sqrt{4({\rho}_1+\Lambda)({\rho}_2-\Lambda)-
(\rho-{\rho}_1-{\rho}_2)^2}-i(\rho-{\rho}_1-{\rho}_2)\right]\right.+\]
\[\sqrt{4({\rho}_1+\Lambda)({\rho}_2-\Lambda)-
(\rho-{\rho}_1-{\rho}_2)^2}\times\]
\[\left.\left.\ln\left[\frac {\sqrt{4({\rho}_1+\Lambda)({\rho}_2-\Lambda)-
(\rho-{\rho}_1-{\rho}_2)^2}-i(\rho-{\rho}_1-{\rho}_2)}
{2i\sqrt{-({\rho}_1+\Lambda)({\rho}_2-\Lambda)}}\right]\right\}\right\}-
\frac {i} {2}\times\]
\[\left\{[\ln(-{\rho}_1+\Lambda)-\ln(-{\rho}_1-\Lambda)]
[\ln(-{\rho}_2+\Lambda)-\ln(-{\rho}_2-\Lambda)]\right.\times\]
\[({\rho}_1-{\rho}_2)\left[\ln\left(i\sqrt{\frac {{\rho}_1+\Lambda}
{{\rho}_2+\Lambda}}\right)+
\ln\left(-i\sqrt{\frac {{\rho}_1-\Lambda}
{{\rho}_2-\Lambda}}\right)\right]+\]
\[[\ln({\rho}_1+\Lambda)-\ln({\rho}_1-\Lambda)]
[\ln({\rho}_2+\Lambda)-\ln({\rho}_2-\Lambda)]\times\]
\[({\rho}_1-{\rho}_2)\left[\ln\left(-i\sqrt{\frac {\Lambda-{\rho}_1}
{\Lambda-{\rho}_2}}\right)+
\ln\left(i\sqrt{\frac {\Lambda+{\rho}_1}
{\Lambda+{\rho}_2}}\right)\right]+\]
\[[\ln({\rho}_1+\Lambda)-\ln({\rho}_1-\Lambda)]
[\ln(-{\rho}_2+\Lambda)-\ln(-{\rho}_2-\Lambda)]\times\]
\[\left\{({\rho}_1-{\rho}_2)\left[\ln\left(\sqrt{\frac {\Lambda+{\rho}_1}
{\Lambda-{\rho}_2}}\right)+
\ln\left(\sqrt{\frac {\Lambda-{\rho}_1}
{\Lambda+{\rho}_2}}\right)\right]\right.+\]
\[\frac {({\rho}_1-{\rho}_2)} {2}\left[\ln(-{\rho}_1-{\rho}_2+\Lambda)-
\ln(-{\rho}_1-{\rho}_2-\Lambda)\right.-\]
\[\left.\ln({\rho}_1+{\rho}_2+\Lambda)+\ln({\rho}_1+{\rho}_2-\Lambda)\right]
+{\rho}_2\left[\ln(-{\rho}_1-{\rho}_2+\Lambda)\right.-\]
\[\left.\left.\ln(-{\rho}_1-{\rho}_2-\Lambda)\right]+{\rho}_1\left[
\ln({\rho}_1+{\rho}_2+\Lambda)-\ln({\rho}_1+{\rho}_2-\Lambda)\right]\right\}\]
\[[\ln(-{\rho}_1+\Lambda)-\ln(-{\rho}_1-\Lambda)]
[\ln({\rho}_2+\Lambda)-\ln({\rho}_2-\Lambda)]\times\]
\[\left\{({\rho}_1-{\rho}_2)\left[\ln\left(\sqrt{\frac {\Lambda-{\rho}_1}
{\Lambda+{\rho}_2}}\right)+
\ln\left(\sqrt{\frac {\Lambda+{\rho}_1}
{\Lambda-{\rho}_2}}\right)\right]\right.+\]
\[\frac {({\rho}_1-{\rho}_2)} {2}\left[\ln({\rho}_1+{\rho}_2+\Lambda)-
\ln({\rho}_1+{\rho}_2-\Lambda)\right.-\]
\[\left.\ln(-{\rho}_1-{\rho}_2+\Lambda)+\ln(-{\rho}_1-{\rho}_2-\Lambda)\right]
+{\rho}_1\left[\ln(-{\rho}_1-{\rho}_2+\Lambda)\right.-\]
\begin{equation}
\label{ep10.1}
\hspace{-6mm}
\left.\left.\left.\left.\ln(-{\rho}_1-{\rho}_2-\Lambda)\right]+{\rho}_2\left[
\ln({\rho}_1+{\rho}_2+\Lambda)-\ln({\rho}_1+{\rho}_2-\Lambda)\right]\right\}
\right\}\right\}\;d{\rho}_1\;d{\rho}_2
\end{equation}
\[|\Im(\rho)|>\Im(\Lambda)>|\Im({\rho}_1)|+|\Im({\rho}_2)|\;;\;
\gamma <\min\left(\frac {\pi} {2\mid\Im(\rho_1)\mid};\frac {\pi} {2\mid\Im(\rho_2)\mid}\right)\]
We define
\begin{equation}
\label{ep10.2}
H(\rho)=H^{(0)}(\rho,i0^+)=H_{\gamma}^{(0)}(\rho,i0^+)
\end{equation}
\begin{equation}
\label{ep10.3}
H_{\gamma\lambda}(\rho,i0^+)=\sum\limits_{-m}^{\infty}
H_{\gamma}^{(n)}(\rho,i0^+){\lambda}^n
\end{equation}
If we take into account that singularities (in the variable $\Lambda$) are
contained in a horizontal band of width $|{\sigma}_0|$ we have:
\begin{equation}
\label{ep10.4}
H_{\gamma\lambda}(\rho,i0^+)=\sum\limits_{-m}^{\infty}
H_{\gamma\lambda}^{(n)}(\rho,i\sigma)\frac {(-i\sigma)^n} {n!}
\;\;\;\;\;\;\sigma>|{\sigma}_0|
\end{equation}
As in the other cases we define now
\begin{equation}
\label{ep10.5}
\{F\ast G\}(\rho)=H(\rho)
\end{equation}
as the convolution of two Lorentz invariant ultradistributions of exponential type.

Let ${\hat{H}}_{\gamma\lambda}(x)$ be the Fourier antitransform of
$H_{\gamma\lambda}(\rho,i0^+)$:
\begin{equation}
\label{ep10.6}
{\hat{H}}_{\gamma\lambda}(x)=\sum\limits_{n=-m}^{\infty}{\hat{H}}_{\gamma}^{(n)}(x){\lambda}^n
\end{equation}
If we define:
\[{\hat{f}}_{\gamma\lambda}(x)={\cal F}^{-1}\{F_{\gamma\lambda}(\rho)\}=
{\cal F}^{-1}\{[\cosh(\gamma\rho)]^{-\lambda}F(\rho)\}\]
\begin{equation}
\label{ep10.7}
{\hat{g}}_{\gamma\lambda}(x)={\cal F}^{-1}\{G_{\gamma\lambda}(\rho)\}=
{\cal F}^{-1}\{[\cosh(\gamma\rho)]^{-\lambda}G(\rho)\}
\end{equation}
then
\begin{equation}
\label{ep10.8}
{\hat{H}}_{\gamma\lambda}(x)=(2\pi)^4{\hat{f}}_{\gamma\lambda}(x){\hat{g}}_{\gamma\lambda}(x)
\end{equation}
and taking into account the Laurent's developments of $\hat{f}$ and
$\hat{g}$:
\[{\hat{f}}_{\gamma\lambda}(x)=\sum\limits_{n=-m_f}^{\infty}{\hat{f}}_{\gamma}^{(n)}(x){\lambda}^n\]
\begin{equation}
\label{ep10.9}
{\hat{g}}_{\gamma\lambda}(x)=\sum\limits_{n=-m_f}^{\infty}{\hat{g}}_{\gamma}^{(n)}(x){\lambda}^n
\end{equation}
we can write:
\begin{equation}
\label{ep10.10}
\sum\limits_{n=-m}^{\infty}{\hat{H}}_{\gamma}^{(n)}(x){\lambda}^n=(2\pi)^4
\sum\limits_{n=-m}^{\infty}\left(\sum\limits_{k=-m}^n {\hat{f}}_{\gamma}^{(k)}(x)
{\hat{g}}_{\gamma}^{(n-k)}(x)\right){\lambda}^n
\end{equation}
$(m=m_f+m_g)$\\
and as a consequence:
\begin{equation}
\label{ep10.11}
{\hat{H}}^{(0)}(x)=
\sum\limits_{k=-m}^0{\hat{f}}_{\gamma}^{(k)}(x){\hat{g}}_{\gamma}^{(n-k)}(x)
\end{equation}

\subsection*{Examples}

As a first example of the use of (\ref{ep10.1}) we shall evaluate the convolution
product of $\delta(\rho)$ with $\delta(\rho-{\mu}^2)$ with $\mu={\mu}_R+i{\mu}_I$
a complex number such that: ${\mu}_R^2>{\mu}_I^2$, ${\mu}_R{\mu}_I>0$.
Thus from (\ref{ep10.1}) we obtain:
\[H_{\gamma\lambda}(\rho,\Lambda)=-i\pi\frac {\ln(-{\mu}^2+\Lambda)-\ln(-{\mu}^2+\lambda)}
{[2\cosh(\gamma{\mu}^2)]^{\lambda}}
\left\{\frac {i(\rho-{\mu}^2)} {8{\pi}^2\rho}\left[
\ln\left(\frac {\rho-{\mu}^2} {\sqrt{\Lambda({\mu}^2+\Lambda)}}\right)+
\right.\right.\]
\[\left.\left.\ln\left(\frac {{\mu}^2-\rho} {\sqrt{-\Lambda({\mu}^2+\Lambda)}}\right)
\right]+\frac {{\mu}^2-\rho} {16\pi\rho}\right\}
-i\pi\frac {\ln(-{\mu}^2+\Lambda)-\ln(-{\mu}^2+\lambda)}
{[2\cosh(\gamma{\mu}^2)]^{\lambda}}\;\;\times\]
\begin{equation}
\label{ep10.12}
\left\{\frac {-i{\mu}^2)} {8{\pi}^2\rho}\left[
\ln\left(\sqrt{\frac {\Lambda} {{\mu}^2+\Lambda}}\right)+
\ln\left(\sqrt{\frac {\Lambda} {\Lambda-{\mu}^2}}\right)
\right]-\frac {{\mu}^2} {16\pi\rho}\right\}
\end{equation}

Simplifying terms and taking the limit $\lambda\rightarrow 0$ (\ref{ep10.15}) turns into:
\[H^{(0)}(\rho,\Lambda)=-i\pi[\ln(-{\mu}^2+\Lambda)-\ln(-{\mu}^2+\lambda)]
\left\{\frac {i(\rho-{\mu}^2)} {8{\pi}^2\rho}\left[
\ln\left(\rho-{\mu}^2\right)+
\right.\right.\]
\begin{equation}
\label{ep10.13}
\left.\left.\ln\left({\mu}^2-{\rho}\right)\right]+
\frac {i{\mu}^2} {8{\pi}^2\rho}\left[\ln({\mu}^2+\Lambda)+
\ln({\mu}^2-\Lambda)\right]\right\}
\end{equation}
Now, if
\[F_1(\mu,\Lambda)=\ln(-{\mu}^2+\Lambda)-\ln(-{\mu}^2-\Lambda)\]
then
\[F_1(\mu,i0^+)=2i\pi\;\;;\;\; {\mu}_R^2>{\mu}_I^2\;\;;\;\;{\mu}_R{\mu}_I>0\]
And, if
\[F_2(\mu,\Lambda)=\ln({\mu}^2+\Lambda)-\ln({\mu}^2-\Lambda)\]
then
\[F_2(\mu,i0^+)=0\;\;;\;\; {\mu}_R^2>{\mu}_I^2\;\;;\;\;{\mu}_R{\mu}_I>0\]
Using these results we obtain:
\begin{equation}
\label{ep10.14}
H(\rho)=\frac {i(\rho-{\mu}^2)} {4\rho}\left[
\ln(\rho-{\mu}^2)+\ln({\mu}^2-\rho)\right]+
\frac {i{\mu}^2} {2\rho}\ln({\mu}^2)
\end{equation}
As a second example we will evaluate the convolution of
$\Theta[\Im(\rho)] e^{ia\rho} $  (a real) with $\delta(\rho)$.

The convolutin can be performed in the real $\rho$-axis to obtain:
\begin{equation}
\label{ep10.15}
h_{\gamma\lambda}(\rho)=\frac {\pi} {2^{\lambda+1}\rho}
\int\limits_{-\infty}^{\infty}\frac {e^{ia{\rho}_2}|\rho-{\rho}_2|}
{[2\cosh(\gamma{\rho}_2)]^{\lambda}}\;d{\rho}_2
\end{equation}
which  can be written as:
\[h_{\gamma\lambda}(\rho)=\frac {\pi} {2^{\lambda+1}}\left[
\frac {i} {\rho} \frac {d} {da}
\int\limits_{-\infty}^{\rho}\frac {e^{ia{\rho}_2}}
{[2\cosh(\gamma{\rho}_2)]^{\lambda}}\;d{\rho}_2+
\int\limits_{-\infty}^{\rho}\frac {e^{ia{\rho}_2}}
{[2\cosh(\gamma{\rho}_2)]^{\lambda}}\;d{\rho}_2\right.-\]
\begin{equation}
\label{ep10.16}
\left.\frac {i} {\rho} \frac {d} {da}
\int\limits_{\rho}^{\infty}\frac {e^{ia{\rho}_2}}
{[2\cosh(\gamma{\rho}_2)]^{\lambda}}\;d{\rho}_2-
\int\limits_{\rho}^{\infty}\frac {e^{ia{\rho}_2}}
{[2\cosh(\gamma{\rho}_2)]^{\lambda}}\;d{\rho}_2\right]
\end{equation}
With the use of the results:
\[\int\limits_{-\infty}^{\rho}\frac {e^{ia{\rho}_2}}
{[2\cosh(\gamma{\rho}_2)]^{\lambda}}\;d{\rho}_2=
\frac { e^{(ia+\gamma\lambda)\rho}} {ia+\gamma\lambda}\;\times\]
\begin{equation}
\label{ep10.17}
F\left(\lambda,\frac {ia+\gamma\lambda} {2\gamma},
\frac {ia+\gamma\lambda} {2\gamma}+1;-e^{-2\gamma\rho}\right)
\end{equation}
\[\int\limits_{\rho}^{\infty}\frac {e^{ia{\rho}_2}}
{[2\cosh(\gamma{\rho}_2)]^{\lambda}}\;d{\rho}_2=
\frac { e^{(ia-\gamma\lambda)\rho}} {\gamma\lambda-ia}\;\times\]
\begin{equation}
 \label{ep10.18}
F\left(\lambda,\frac {\gamma\lambda-ia} {2\gamma},
\frac {\gamma\lambda-ia} {2\gamma}+1;-e^{2\gamma\rho}\right)
\end{equation}
in the limit $\lambda\rightarrow 0$ we obtain:
\begin{equation}
\label{ep10.19}
h(\rho)=-\frac {\pi} {a^2}\frac {e^{ia\rho}} {\rho}
\end{equation}
and therefore, in the complex $\rho$-plane, the corresponding
ultradistribution of exponential type is:
\begin{equation}
\label{ep10.20}
H(\rho)=-\frac {\pi} {a^2\rho}\left\{\Theta[\Im(\rho)]
e^{ia\rho}-\frac {1} {2}\right\}
\end{equation}
As final example we evaluate the convolution between
$F(\rho)=(1/2)Sgn[\Im(\rho)]$ $e^{ia\rho} \cosh({\rho}^{1/2})$ (a real)
and $G(\rho)=\delta(\rho)$ . We perform the calculation of the convolution
in the real $\rho$-plane and then we pass tho the complex
$\rho$-plane.
By the use of the Taylor's development of $\cosh({\rho}^{1/2})$
\begin{equation}
 \label{ep10.21}
\cosh({\rho}^{1/2})=\sum\limits_{n=0}^{\infty}
\frac {{\rho}^n} {2n!}
\end{equation}
we obtain
\[h_{\gamma\lambda}(\rho)=\frac {\pi} {2\rho}\sum\limits_{n=0}^{\infty}
 \frac {(-i)^n} {2n!}\frac {{\partial}^n} {{\partial}a^n}
\int\limits_{-\infty}^{\infty}e^{ia{\rho}_1}\delta({\rho}_2) \; \times\]
\[\frac {[(\rho-{\rho}_1-{\rho}_2)^2-4{\rho}_1{\rho}_2]_+^{\frac {1} {2}}}
{[\cosh(\gamma{\rho}_1)]^{\lambda}[\cosh(\gamma{\rho}_2)]^{\lambda}}
\;d{\rho}_1\;d{\rho}_2\; =\]
\begin{equation}
\label{ep10.22}
\frac {\pi} {2\rho}\sum\limits_{n=0}^{\infty}
\frac {(-i)^n} {2n!}\frac {{\partial}^n} {{\partial}a^n}
\int\limits_{-\infty}^{\infty}
\frac {e^{ia{\rho}_1}|\rho-{\rho}_1|} {[\cosh(\gamma{\rho}_1)]^{\lambda}}
\; d{\rho}_1
\end{equation}
By means of the use of equations (\ref{ep10.17}), (\ref{ep10.18})
and in the limit $\lambda\rightarrow 0$ we obtain:
\begin{equation}
\label{ep10.23}
h(\rho)=-\pi\left(1+\frac {i} {\rho} \frac {\partial} {\partial a}\right)
\sum\limits_{n=0}^{\infty}\frac {(-i)^n} {2n!}
\frac {{\partial}^n} {\partial a^n}
\left(\frac {e^{ia\rho}} {a}\right)
\end{equation}
and consequently:
\[H(\rho)=\pi\left[\left(\frac {\Theta[\Im(\rho)]} {\rho} \frac {\partial} {\partial a}
-\frac {i} {2}Sgn[\Im(\rho)]\right)
\sum\limits_{n=0}^{\infty}\frac {(-i)^n} {2n!}
\frac {{\partial}^n} {\partial a^n}
\left(\frac {e^{ia\rho}} {a}\right)\right]\; +\]
\begin{equation}
\label{ep10.24}
\frac {\pi} {2\rho}\sum\limits_{n=0}^{\infty}
\frac {i^n} {2n!} \frac {(n+1)!} {a^{n+2}}
\end{equation}
As an example of the use of (\ref{ep10.11}) we will evaluate the convolution
product of two Dirac's delta: $\delta(\rho)\ast\delta(\rho)$. In this case we have:
\begin{equation}
\label{ep10.25}
F_{\gamma\lambda}(\rho)=-\frac {[\cosh(\gamma\rho)]^{\lambda}} {2\pi i\rho}=
-\frac {1} {2\pi i \rho}
\end{equation}
and as a consequence:
\begin{equation}
\label{ep10.26}
f_{\gamma\lambda}(\rho)=\delta(\rho)
\end{equation}
The Fourier antitransform of (\ref{ep10.26}) is:
\begin{equation}
\label{ep10.27}
{\hat{f}}_{\gamma\lambda}(x)=\frac {2} {(2\pi)^3}x^{-1}
\end{equation}
Thus we have:
\begin{equation}
\label{ep10.28}
{\hat{f}}_{\gamma\lambda}^2(x)=\frac {4} {(2\pi)^6} x^{-2}
\end{equation}
From (\ref{ep10.28}) we obtain:
\begin{equation}
\label{ep10.29}
\lim_{\lambda\rightarrow 0}{\hat{f}}_{\gamma\lambda}^2(x)=
\frac {4} {(2\pi)^6}x^{-2}
\end{equation}
and taking into account that:
\begin{equation}
\label{ep10.30}
{\cal F}\{x^{-2}\}=\frac {{\pi}^3} {2} Sgn(\rho)
\end{equation}
we obtain
\begin{equation}
\label{ep10.31}
\delta(\rho)\ast\delta(\rho)=\frac {\pi} {2} Sgn(\rho)
\end{equation}

\newpage

\section{Discussion}

In a first paper \cite{tp3} we have shown the existence of the convolution
of two one-dimensional tempered ultradistributions. In a second paper
ref.\cite{tp4} we have
extended these procedure to n-dimensional space. In four-dimensional
space we have given a expression for the convolution of two tempered
ultradistributions (even) in the variables $k^0$ and $\rho$.
In a third paper \cite{tp5}we have defined spherically symmetric and
Lorentz invariant tempered ultradistributions and we have given
the formulaes for the Fourier transform and
four-dimensional convolution of them.

In this paper we have extended these results to ultradistributions
of exponential type and,
in a intermadiate step of deduction we have obtained the generalization
to ultradistributions of exponential type in
Euclidean and Minkowskian space of dimensional regularization in
configuration space (ref.\cite{tp5,tp8})

Furthermore, as an application of our formalism
we have solved the question
of normalization of Gamow States and we have given an example .

When we use the perturbative development in Quantum Field Theory, we
have to deal with products of distributions in configuration space,
or else, with convolutions in the Fourier transformed p-space.
Unfortunately, products or convolutions ( of distributions ) are
in general ill-defined quantities.
On the contrary, the convolution of ultradistributions
is a well-defined quantity and the correspondig
product in configuration space is a true product in a
ring with zero-factors.

In physical applications
one introduces some ``regularization'' scheme, which allows us to
give sense to divergent integrals. Among these procedures we would
like to mention the dimensional regularization method ( ref.
\cite{tp12,tp13} ). Essentially, the method consists in the
separation of the volume element ( $d^{\nu}p$ ) into an angular
factor ( $d\Omega$ ) and a radial factor ( $p^{\nu-1}dp$ ).
First the angular integration is carried out and then the number
of dimensions $\nu$ is taken as a free parameter. It can be adjusted
to give a convergent integral, which is an analytic function of
$\nu$.

Our formula (\ref{ep10.1}) is similar to the expression one obtains with
dimensional regularization. However, the parameter $\lambda$ is
completely independents of any dimensional interpretation.

All ultradistributions provide integrands
(in (\ref{ep6.1}),(\ref{ep6.27}),(\ref{ep8.1}),(\ref{ep9.1}),(\ref{ep10.1})) that are
analytic functions along the integration path. The lambda parameters
permit us to control the possible exponential asymptotic
behavior ( cf. eq.(\ref{er2.25})). The existence of a region of
analyticity for  $\lambda$, and a subsequent continuation to
the point of interest, defines the convolution
product.

The properties described below
show that ultradistributions provide an
appropriate framework for applications to physics. Furthermore,
they can ``absorb'' arbitrary pseudo-polynomials
(in the variables $s_i=e^{z_i}$), thanks to eq.(\ref{er2.16}).
A property that is interesting for renormalization theory.
 For this reason we decided to begin this paper (and also for the benefit
 of the reader), with a summary of the main characteristics
of ultradistributions of exponential type and their Fourier
transform.

As a final remark we would like to point out that our formulae
for convolutions are general definitions and not regularization methods.

\end{document}